\documentclass[12pt,a4paper]{article}

\pdfoutput=1
\usepackage{jheppub}

\usepackage{bm}
\usepackage{ifsym}
\usepackage{amsthm}
\usepackage{amsfonts}
\usepackage{ccaption}
\usepackage{mathrsfs}
\usepackage{booktabs}

\makeatletter
\@addtoreset{equation}{section}

\makeatletter
\renewcommand\section{\@startsection {section}{1}{\z@}%
                                   {-3.5ex \@plus -1ex \@minus -.2ex}
                                   {2.3ex \@plus.2ex}%
                                   {\normalfont\large\bfseries}}
\renewcommand\subsection{\@startsection{subsection}{2}{\z@}%
                                     {-3.25ex\@plus -1ex \@minus -.2ex}%
                                     {1.5ex \@plus .2ex}%
                                     {\normalfont\bfseries}}

  \captionnamefont{\bfseries}
  \captiontitlefont{\small\sffamily}
  \captiondelim{: }
  \hangcaption

\def\sec#1{\S\ref{#1}}
\def\fig#1{Fig.\,\ref{#1}}
\def\req#1{(\ref{#1})}

\def\thuss{\qquad \Longrightarrow \qquad}

\def\ands{\qquad {\rm and} \qquad}

\def\RR{{\mathbf R}}

\def\ket#1{\mid  \! #1  \rangle}

\definecolor{rust}{rgb}{0.8,0.2,0.2}
\definecolor{green}{rgb}{0.1,0.8,0.2}

\def\MR#1{{\color{blue}{ #1}}}

\def\AdS#1{AdS$_{#1}$}

\def\bulk{{\cal M}}
\def\bdy{{\cal B}}
\def\cbulk{{\bar {\cal M}}}

\def\curt#1{\gamma_{#1}}
\def\curf#1{\gamma_{#1}^+}
\def\curp#1{\gamma_{#1}^-}
\def\set#1{\{ \, #1 \, \}}
\def\st{ \, \mid \, }

\def\ro{r_0}

\def\entsurf{\partial {\cal A}}
\def\caustic{{\mathscr C}}
\def\cwedge{\blacklozenge_{{\cal A}}}

\def\bcwedge{\partial_{{\cal M}}(\blacklozenge_{{\cal A}})}
\def\bcwedgef{\partial_+(\blacklozenge_{{\cal A}})}
\def\bcwedgep{\partial_-(\blacklozenge_{{\cal A}})}

\def\domd{\Diamond_{\cal A}}
\def\domdc{\Diamond_{{\cal A}^c}}

\def\csfz#1{{ \Psi}_{#1}}
\def\extr#1{{\mathfrak E}_{#1}}
\def\csf#1{{\Xi}_{#1}}

\def\domdy{D_{\bdy}} 
\def\domiy{J_{\bdy}}  
\def\ddtipp{{q^\vee}}  
\def\ddtipf{{q^\wedge}}  

\def\ro{{r_0}}

\title{Causal Holographic Information}

\author{Veronika E. Hubeny}
\author{ \& Mukund Rangamani}

\affiliation[a]{ Centre for Particle Theory \& Department of Mathematical Sciences,\\
Science Laboratories, South Road, Durham DH1 3LE, UK.}
\affiliation[b]{Kavli Institute for Theoretical Physics, University of California,\\
Santa Barbara, CA 93015, USA.}
\emailAdd{veronika.hubeny@durham.ac.uk}
\emailAdd{mukund.rangamani@durham.ac.uk}

\abstract{We propose a measure of holographic information based on a causal wedge construction. The motivation behind this comes from an attempt to understand how boundary field theories can holographically reconstruct spacetime. We argue that given the knowledge of the reduced density matrix in a spatial region of the boundary, one should be able to reconstruct at least the corresponding bulk causal wedge. 
In attempt to quantify the `amount of information' contained in a given spatial region in field theory, we consider a particular bulk surface (specifically a co-dimension two surface in the bulk spacetime which is an extremal surface on the boundary of the bulk causal wedge), and propose that the area of this surface, measured in Planck units, naturally quantifies the information content.  
We therefore call this area the causal holographic information. We also contrast our ideas with earlier studies of holographic entanglement entropy. In particular, we establish that the causal holographic information, whilst not being a von Neumann entropy, curiously enough agrees with the entanglement entropy in all cases where one has a microscopic understanding of entanglement entropy.} 


\begin{document}
\begin{flushright} \small{DCPT-12/21} \end{flushright}

\maketitle

\flushbottom



\section{Introduction}
\label{s:intro}

Holography is a curious thing: while on the one hand it makes it clear that non-gravitational quantum dynamics can in certain circumstances be reformulated as a classical gravitational system, it obscures the precise manner of this reincarnation of the degrees of freedom. Various attempts have been made in the past to pierce this veil of holographic mystique, a natural question to focus on in this context is locality in the gravitational description. A-priori the mechanics of the gauge/gravity correspondence suggests that the entire field theory path integral on the whole maps to the bulk geometry. It is however interesting to ask how much of the bulk geometry is `known' to a given part of the boundary field theory. Motivated by this question we are led to a new measure of holographic ``information", which we dub {\em  causal holographic information}, $\chi$.

To motivate the discussion, let us step back and consider a  quantum field theory defined on a $d$-dimensional Lorentzian background ${\cal B}_d$ with a prescribed non-dynamical metric $\gamma_{\mu\nu}$.  This background is rigid, and we will for simplicity assume that it admits a foliation by spacelike Cauchy slices $\Sigma_{\cal B}$ on which we can prescribe initial data and subsequently evolve. The  holographic picture asserts that given the path integral on the background ${\cal B}_d$, we can find an asymptotically locally AdS geometry $\bulk_{d+1}$ whose boundary $\partial {\cal M}_d$ is in the conformal class of $\bdy_d$ and we will typically conflate these $d$ dimensional geometries. Classical gravitational dynamics on $\bulk_{d+1}$ has as its hologram the entire path integral (in the planar limit) of the dual field theory.

Now let us consider a specific Cauchy slice $\Sigma_{\cal B}$ and focus on a particular spatial region ${\cal A} \in \Sigma_{\cal B}$ and ask: what part of the bulk is ${\cal A}$ supposed to be cognizant of? Even though we believe the holographic reconstruction of spacetime is non-local, it nevertheless seems implausible that the specific region ${\cal A}$ would be aware of the entire bulk geometry.\footnote{
One useful point to keep in mind is that the field theory observables could be used to probe a region of the bulk spacetime which is larger than the region that can be reconstructed using the data in ${\cal A}$. Moreover, we also remark here that we will refrain from assuming any analyticity property of the bulk geometry. For if one were to do so, then one could trivially reconstruct the entire bulk using the knowledge of an arbitrarily small open neighborhood.} Rather, it seems natural that there is information regarding a certain subset of the bulk that is contained in quantum operators localized to within ${\cal A}$ (or at least within the boundary region whose physics is fully determined by  ${\cal A}$); we want to quantify what this measure is.  Instead of building up this notion directly, which would require specifying a plausible reconstruction of the bulk, we proceed in reverse.  We first  observe that certain bulk constructs are very `natural', and then try to understand them from the field theory point of view. 

Let us start by asking the following bulk question: given some asymptotically-AdS bulk spacetime, and a specific spatial region ${\cal A}$ on the boundary, what is the `most natural' bulk region that one can associate to ${\cal A}$?  The answer of course depends on what we mean by natural.  We could take the viewpoint that in a Lorentzian spacetime, causal relations are in some sense more basic and therefore natural than the geometry.  But even without explicitly elevating the causal structure above geometry, it is clear that causal structure is an important characteristic of the spacetime.   Not only is it fully covariant, but it is also invariant under conformal transformations.  It is therefore tempting to define a `natural region' associated to ${\cal A}$ purely by causal relations.
In particular, we can rephrase our question as: what is the minimal non-trivial bulk (co-dimension 0) region associated to ${\cal A}$ purely by causal relations?
   We now argue that such a region is the causal wedge, $\cwedge$.
 
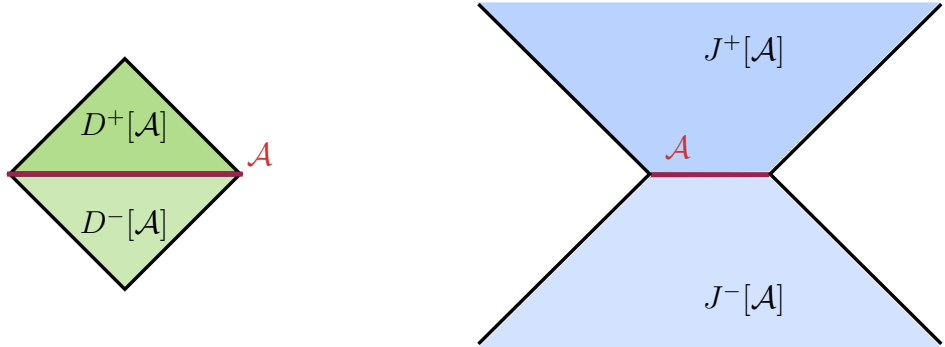
\begin{figure}
\begin{center}
\includegraphics[width=5in]{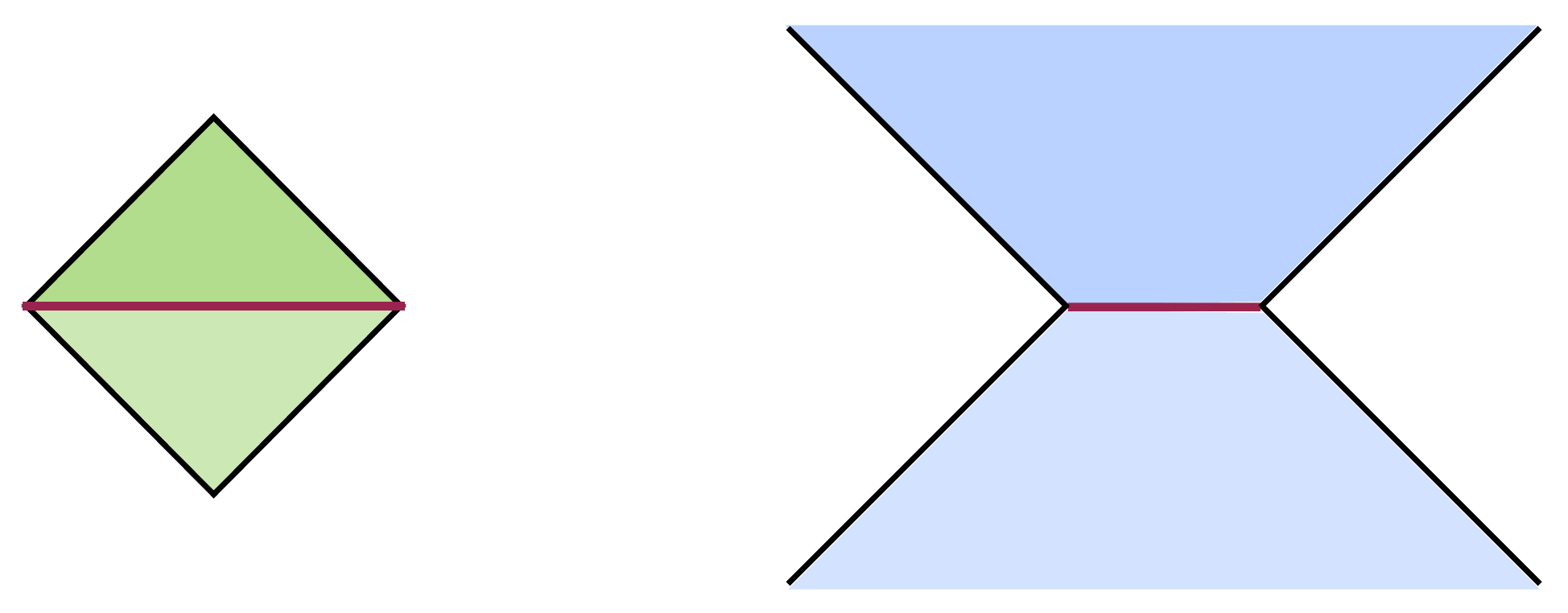}
\begin{picture}(0,0)
\setlength{\unitlength}{1cm}
\put(-11.7, 3){$D^+[{\cal A}]$}
\put(-11.7, 1.7){$D^-[{\cal A}]$}
\put(-3.5, 4){$J^+[{\cal A}]$}
\put(-3.5, .7){$J^-[{\cal A}]$}
\put(-4, 2.7){{\color{rust}{${\cal A}$}}}
\put(-9.5, 2.6){{\color{rust}{${\cal A}$}}}
\end{picture}
\caption{
Illustration of the causal sets $D$ and $J$ associated with a 1-dimensional spacelike region ${\cal A}$.  The future (past) domain of {\it dependence} $D^{\pm}[{\cal A}]$ is the set of points which are fully determined by future (past) evolution of the `initial data'  on ${\cal A}$.  The   future (past) domain of {\it influence} $J^{\pm}[{\cal A}]$ is the set of points which can be causally influenced by (or can influence) ${\cal A}$.
}
\label{f:csets}
\end{center}
\end{figure}

The most natural causal sets pertaining to a given region are the domain of dependence $D$ and the domain of influence $J$, indicated in \fig{f:csets}.   As the terminology suggests, domain of influence  $J$ of some region is the set of all spacetime points which {\it can} be causally influenced by or influence events in that region, while the domain of dependence $D$ is the much smaller set of spacetime points which {\it must} be causally influenced by or influence events in that region.   We distinguish future and past domains of the region by whether they lie to the future or past of that region.  Moreover, one can consider such sets to be confined purely to the boundary, or alternately to pertain to the full bulk.  These are all the ingredients needed for specifying the region of interest, $\cwedge$.  We define these notions  precisely in \sec{s:construct}; for now we simply motivate  the construction.

How can we construct  a minimal $d+1$ dimensional bulk region from a $d-1$-dimensional spatial region ${\cal A}$ on the boundary?  Clearly, both bulk and boundary domains of influence of ${\cal A}$ are infinite sets.  Their union is likewise infinite, while their intersection is just the region ${\cal A}$ itself, so none of these provides a good starting point.  On the other hand, the bulk domain of dependence of ${\cal A}$ is only the region ${\cal A}$ itself, which doesn't extend into the bulk.\footnote{
As follows immediately from the definition we give in \sec{s:construct}, the domain of dependence of a given region trivializes to just the region whenever that region has co-dimension greater than 1.
} This leaves us with the boundary domain of dependence of ${\cal A}$, which we'll denote by $\domd$.  This is a finite $d$-dimensional region on the boundary, and we will use this boundary region to construct the bulk region of interest $\cwedge$.  While the bulk domain of dependence of $\domd$ is still only $\domd$, and the bulk domain of influence of $\domd$ is still infinitely extended, the {\it intersection} of future and past domains of influence of $\domd$ is now a non-trivial bulk region which  nevertheless does not extend infinitely far into the bulk.  This is our region $\cwedge$, called the {\it causal wedge} of $\domd$.  For orientation we refer the reader to \fig{f:cwedge} of the next section, where we explain the technical construction.

Having constructed a $d+1$ dimensional bulk region $\cwedge$ associated with a $d-1$ dimensional boundary region ${\cal A}$, let us go one step further, and ask whether there is likewise a  $d-1$ dimensional bulk `surface' naturally associated to ${\cal A}$, as this may provide a more useful (albeit more limited)  quantity  related to ${\cal A}$.  We can again answer in the affirmative, by building on the construction of $\cwedge$:  keeping to only causally-defined quantities, we define the surface of interest $\csf{\cal A}$ as the  (bulk) intersection of the boundaries of the past and future domains of influence of $\domd$.  The boundaries are null surfaces in the bulk which end on the boundary of $\domd$, so their intersection is a spacelike co-dimension 2 bulk surface which  is anchored on the AdS boundary at $\partial {\cal A}$, and for static geometries lies entirely within the same time slice\footnote{
For static bulk geometries, we use the natural time slices defined by the time translation symmetry, i.e.\ orthogonal to the timelike Killing field.} as ${\cal A}$.
More generally, $\csf{\cal A}$  corresponds to a surface within $\cwedge$ which reaches deepest into the bulk.\footnote{
In fact, among the restricted set of surfaces anchored on $\partial {\cal A}$ and lying entirely within the boundary of $\cwedge$, $\csf{\cal A}$ an extremal surface. }
  We will call $\csf{\cal A}$ the {\it causal information surface}  for reasons we indicate below.  

So far, we have used only causal relations to single out two bulk quantities naturally associated to the boundary region ${\cal A}$, namely the causal wedge $\cwedge$ and the causal information surface $\csf{\cal A}$. Let us now include the information contained in the bulk geometry, and try to distill some minimal non-trivial information about these bulk regions which might characterize ${\cal A}$.  The most natural quantities which pertain to $\cwedge$ and $\csf{\cal A}$ are the proper spacetime volume and proper area, respectively.  Since both bulk regions extend out to the boundary, both of these quantities are a-priori infinite; however we can usefully consider how the divergence depends on ${\cal A}$, and moreover there is a natural way to regulate them (see for e.g., \cite{Ryu:2006bv,Ryu:2006ef}).  Since ${\cal A}$ and $\csf{\cal A}$ have the same dimensionality, we will focus on the quantity corresponding to the proper area of $\csf{\cal A}$; we will dub this {\it causal holographic information} and denote it by $\chi_{{\cal A}}$.  

We have now motivated all the bulk quantities we wish to consider.  To summarize, we have argued, using only causality, that the causal wedge $\cwedge$ is the smallest non-trivial $d+1$-dimensional bulk region specified by ${\cal A}$, while a related $d-1$ dimensional surface most naturally associated with ${\cal A}$ is the causal information surface $\csf{{\cal A}}$, which has proper area $\chi_{{\cal A}}$.  Given the simple and fundamental nature of the construction from the bulk point of view, the main question we wish to pose is: {\it What is the meaning of the bulk region $\cwedge$ and the number $\chi_{{\cal A}}$ from the point of view of the CFT?}

There are two complementary reasons to consider this question, one motivated from the gravity side and one from the CFT side.
From the bulk standpoint, understanding what kind of CFT quantity corresponds to $\cwedge$ would give us crucial insight into how bulk causality is encoded in the dual field theory, which may in turn elucidate the emergence of dynamical spacetime from the field theory.   On the other hand, if $\chi_{{\cal A}}$ characterizes some useful CFT quantity, calculating it by using the bulk construction may be the most efficient way to study it.  

Let us then turn to the field theory side.  What are the `natural' quantities associated to a region ${\cal A}$ from the boundary standpoint?  One set of quantities which immediately springs to mind is the entanglement entropy  $ S_{\cal A}$ which is the von Neumann entropy, or more generally the Renyi entropies $S_{\cal A}^{(n)}$ \cite{Renyi:1960fk},  of the reduced density matrix $\rho_{\cal A}$  
\begin{equation}
S_{\cal A} = -{\rm Tr}(\rho_{\cal A} \, \log \rho_{\cal A}) \ ; \qquad S_{\cal A}^{(n)} = \frac{1}{1-n} \, \log\left[ {\rm Tr}(\rho_{\cal A}^n) \right] \ ,
\label{eedef}
\end{equation}	
with the latter encapsulating the moments of the density matrix $\rho_{\cal A}$. The reduced density matrix is obtained by carrying out the path integral over fields supported in the complement $\Sigma_{\cal B}\backslash{\cal A}$ of the region ${\cal A}$.  Let us take a brief detour to remind the reader of the significance of these quantities.

Entanglement entropy and Renyi entropies have been of much interest in quantum information and quantum field theory literature and they have been studied extensively in the holographic context following the remarkable proposal of Ryu and Takayanagi  to compute the holographic entanglement entropy \cite{Ryu:2006bv,Ryu:2006ef}. This proposal which is remarkable in  its simplicity, associates $S_{\cal A}$ to the area of a bulk minimal spacelike co-dimension two surface $\extr{\cal A}$ (at constant time) anchored on the boundary
$\partial {\cal A}$ of the desired region ${\cal A}$. 
For states that are not static, we need to relax the minimality constraint and work rather with extremal surfaces as originally  described in \cite{Hubeny:2007xt}. This latter construction also allows one to make contact between covariant entropy bounds in gravitational theories captured by light-sheets \cite{Bousso:2002ju} and entanglement entropy.

While this proposal has not been derived from first principles in general, there is mounting evidence that it is correct. The first attempt to derive the minimal surface proposal was made in \cite{Fursaev:2006ih}, which was critically examined in \cite{Headrick:2010zt}, who pointed out some loopholes in the above derivation.\footnote{
In particular the derivation of \cite{Fursaev:2006ih} proceeds by using the replica trick to relate the computation of entanglement entropy to computation of a partition function on a multi-sheeted geometry built from $\Sigma_{\cal B}\backslash{\cal A}$ which is evaluated in turn using saddle point techniques. The issue pointed out by \cite{Headrick:2010zt} is that the putative saddle point does not correspond to a true solution to all equations of motion and in particular fails to reproduce known results for Renyi entropies.} More recently in a beautiful analysis \cite{Casini:2011kv} demonstrated that the holographic minimal surface prescription can indeed be derived for spherical entangling regions in flat space i.e., for  $\partial{\cal A} = {\bf S}^{d-2} \subset {\mathbb R}^{d-1} = \Sigma_{\cal B}$ by using a conformal transformation to map the reduced density matrix to a thermal density matrix and finally computing the latter using black hole entropy as is standard in AdS/CFT (Renyi entropies were computed using this trick in \cite{Hung:2011nu}). 

Based on the discussion above it would seem that we've identified the object of interest: couldn't the causal holographic information  $\chi_{\cal A}$ associated with ${\cal A}$ be just identified with the corresponding entanglement entropy $S_{\cal A}$?\footnote{
The idea of using entanglement in field theory to understand how a holographically dual spacetime can emerge has been discussed in 
\cite{VanRaamsdonk:2009ar,VanRaamsdonk:2010pw,VanRaamsdonk:2011zz}.} 
This is more subtle than one might naively imagine; indeed we will argue that the answer is ``not in general'': as we will see in explicit examples, the causal information surface $\csf{{\cal A}}$ does {\it not} in general coincide with the extremal surface $\extr{\cal A}$, and correspondingly the causal holographic information  $\chi_{\cal A}$ is  {\it not}  equivalent to the entanglement entropy $S_{\cal A}$. 

However, even though these quantities do not coincide in general, there are specific situations in which they {\it do} agree.  Rather remarkably, such situations are precisely the ones in which we can actually compute the entanglement entropy directly in the CFT! In particular, {\it  in all cases where one is able to compute entanglement entropy in quantum field theories from first principles independently of the strength of quantum interactions, the surfaces $\extr{\cal A}$ and $\csf{\cal A}$ agree}. We believe that this curious agreement is more than mere coincidence, and points to an underlying relation between these quantities.  Moreover, as we will argue below, our causal construction provides a  certain bound on the entanglement entropy.

Emboldened by this observation, we can speculate why does $\chi_{\cal A}$ in some sense characterize the `information content' in ${\cal A}$.
Since the causal wedge $\cwedge$ corresponds to the set of bulk points which can both influence and be influenced by the boundary region $\domd$ which is determined solely by ${\cal A}$, it is tempting to propose that the bulk region $\cwedge$ can be reconstructed from ${\cal A}$ more easily than the rest of the bulk geometry.  For example, one might imagine that one `creates' a local bulk observer by acting with a local CFT operator in the early part of $\domd$.  Having thus a field theory `handle' on such an observer, one lets that observer explore the bulk spacetime and eventually come back to the boundary, having collected information about the traversed region of the bulk.  If one restricts that the observer return to the boundary within $\domd$, the amount of spacetime that may be thusly probed is precisely the causal wedge $\cwedge$. 

How many independent `bits' of information can an ensemble of such observers collect?  Naively, one might imagine that, by bulk locality, every spacetime point is independent, so that the total information should scale with the spacetime volume.  However, this is not correct in a gravitational theory, which is famously governed by the holographic principle \cite{tHooft:1993gx,Susskind:1994vu}. Moreover, although $\cwedge$ extends in time, if we know the bulk field equations, we could reduce the information in $\cwedge$ to just initial data on a spacelike slice in the bulk extending between the AdS boundary and $\csf{{\cal A}}$, since for known boundary conditions on AdS boundary, $\cwedge$ is in fact the bulk domain of dependence for such an initial data slice.   (After all, we are already using the CFT evolution on the boundary in our starting assumption that the physical data specified on ${\cal A}$ automatically determines physics in the full boundary domain of dependence $\domd$.)
Combining these two reductions, one is left with just ${\cal A}$ and the causal information surface $\csf{{\cal A}}$.  It is then natural to ascribe the information contained in ${\cal A}$ to the area of $\csf{{\cal A}}$ in Planck units, namely $\chi_{\cal A}$.

A related observation which one can make is the following:  consider the computation of  correlation functions for gauge invariant boundary operators inserted within $\domd$ in Lorentzian AdS/CFT. Since these correlation functions can be obtained by taking the boundary limit of appropriate bulk correlation functions \cite{Banks:1998dd} (see also \cite{Harlow:2011ke} for recent discussions), and  since the latter are sensitive to the bulk causal structure, it follows that the information contained in the causal wedge should be sufficient to compute the boundary correlators of interest. This point has been made precise in the detailed analysis of \cite{Marolf:2004fy}, which shows that the boundary correlation functions may be determined solely by the bulk path integral over $\cwedge$.

Having motivated  the surface $\csf{\cal A}$ and its associated information measure $\chi_{\cal A}$, we proceed to show that in general this measure of information is however {\it not} a von Neumann entropy. In particular, it fails in certain calculable circumstances to satisfy an important convexity requirement known as strong sub-additivity. Nevertheless, we argue that $\chi_{\cal A}$ provides an upper bound on the entanglement entropy. We note in passing that  in fact, the Renyi entropies introduced in \eqref{eedef} also fail to be sub-additive.

One way to motivate the causal wedge construction and understand the coincidence of the holographic information surface $\csf{{\cal A}}$ with the extremal surface 
$\extr{\cal A}$
 in suitably symmetric regions, is to note that such symmetric regions (which we review in \sec{s:connections}) are the ones which allow maximal entanglement between the degrees of freedom localized in ${\cal A}$ and its complement.  One piece of supporting evidence in favor of this notion comes from the recent investigations of \cite{Hubeny:2012ry}, who showed that the reach of the extremal surfaces is maximal for such symmetric regions.\footnote{
The statement needs a qualifier: at fixed area of the boundary region ${\cal A}$, the shape of ${\cal A}$ which maximizes the reach of the corresponding extremal surface into the bulk is the round ball.  In \cite{Hubeny:2012ry} this was shown for static planar geometries, but we expect it to remain true for sufficiently slow variations of the bulk geometry.} 
For special states such as the vacuum, one can make a stronger statement.
From a field theoretic standpoint one can understand this by noting that the density matrices associated with such regions can be converted to a thermal density matrix by a conformal transformation as for example discussed in  \cite{Casini:2011kv}. This suggests that in a suitable conformal frame the degrees of freedom in the region ${\cal A}$ under consideration are indeed maximally entangled with the rest of the field theoretic degrees of freedom. 
This is related to the fact that black hole entropy encodes the maximal amount of information one can pack in a given region. In terms of information content, they should then enable us to reconstruct the maximal amount of the bulk spacetime.\footnote{
We would like to thank Don Marolf for  insightful comments on this issue.} 

Some of the above observations were already made in \cite{Hubeny:2007xt} where the construction of the surface $\csf{\cal A}$ was proposed and investigated as a potential candidate for a covariant formulation of entanglement entropy. It was also noted there that for certain special cases the result for $\chi_{\cal A}$ disagrees with the result for entanglement entropies extrapolated from free field theories, while the minimal (or more generally extremal) surfaces interpolate more cleanly from the weak coupling result to the holographic domain of strong coupling, and further speculated that $\chi_{\cal A}$ bounds the entanglement from above.

The organization of this paper is as follows: we begin in \sec{s:construct} by presenting the basic causal construction of the surfaces and regions of interest and further lay out our proposal for the causal holographic information. We then examine the circumstances where  $\chi_{\cal A}$ and $S_{\cal A}$ agree in \sec{s:connections} and explain in what manner might $\chi_{\cal A}$ correspond to the maximal information contained in a local region of the field theory. We  conclude with a discussion of various aspects of our proposal and open questions in \sec{s:discuss}. As our discussion involves various regions in both the bulk and the boundary we collect the definitions of these in a Table \ref{t:defns} for quick reference.

\paragraph{Note added:} While this manuscript was under preparation, the articles \cite{Bousso:2012sj}  and \cite{Czech:2012bh} appeared on the arXiv which have some overlap with our considerations. In \cite{Bousso:2012sj} the authors argue that the region of the bulk that should be holographically described by a boundary spacetime region is given by the causal construction based on covariant entropy bounds and light-sheet constructions (for the boundary region). On the other hand \cite{Czech:2012bh} consider various constraints on the region of the bulk that can be constructed from the field theory density matrix and suggest that such a bulk region could be larger than the causal wedge. We thank Mark van Raamsdonk for sharing the results of \cite{Czech:2012bh} with us prior to publication and for many useful discussions.

\section{The construction}
\label{s:construct}

\begin{figure}
\begin{center}
\includegraphics[width=2.5in]{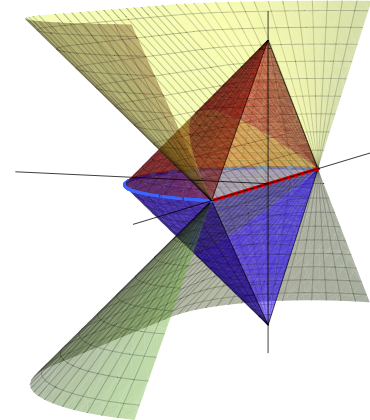}
\begin{picture}(0,0)
\setlength{\unitlength}{1cm}
\put(-5.4,3.4){{\vector(2,1){1}}}
\put(-6,3.1){{\color{blue}{$\csf{{\cal A}}$}}}
\put(-6.5, 4.5){$z$}
\put(-1.8,7.2){$t$}
\put(0, 4.8){$x$}
\put(-2.3, 3.6){{\color{rust}{${\cal A}$}}}
\put(-3.8, 6.7){$J^+[{\cal A}]$}
\put(-3.8, .7){$J^-[{\cal A}]$}
\put(-2.5, 4.6){{\color{black}{$\cwedge$}}}
\end{picture}
\caption{
Illustration of various causal sets associated with the boundary region ${\cal A}$ (color online). AdS boundary is the plane at $z=0$ on the right; the bulk extends to the left of this plane.  The region ${\cal A}$ is the red segment at $z=0, t=0$.  
The future and past bulk domains of influence of ${\cal A}$  are bounded by yellow and green surfaces respectively, and future and past boundaries of the bulk causal wedge $\cwedge$ are indicated by the red and blue surfaces respectively.  Their intersection with the AdS boundary encloses the boundary domain of dependence $\domd$, and their intersection with each other (light-blue curve) corresponds to the causal information surface $\csf{{\cal A}}$.
For simplicity we illustrate these constructs in Poincare AdS; the causal wedge in global AdS$_3$ appears in \fig{f:cwedge3d}(a) (which shows ${\cal A}$ corresponding to half the circle; causal wedge of any other interval would be obtained simply by translating one of the null planes with respect to the other). 
}
\label{f:cwedge}
\end{center}
\end{figure}
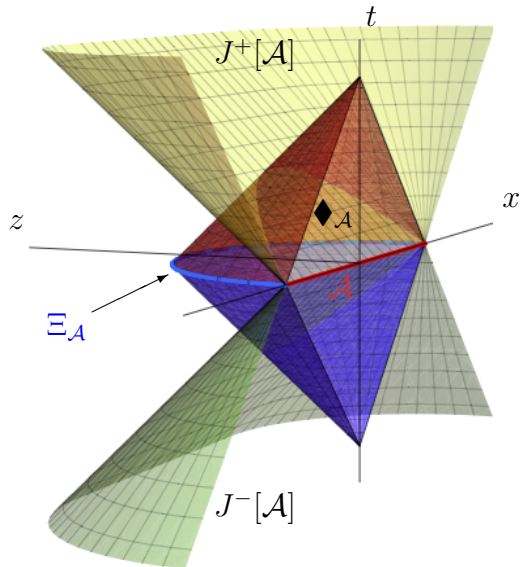

We begin in this section by outlining the basic construction of the causally motivated surface $\csf{\cal A}$. Readers who are familiar with the notions of domains of dependence and causal wedges might prefer to skip directly to \sec{s:connections} where we describe some of the properties of the construction, consulting \fig{f:cwedge} or Table \ref{t:defns} for our notation.

%

\subsection{The holographic causal surface}
\label{s:hcausal}

Consider a $d+1$-dimensional, asymptotically locally AdS spacetime, which we take to be causally well-behaved. We will refer to this bulk geometry as $\bulk$ and its timelike boundary as $\partial{\cal M}$, taken to be in the conformal class of a fixed background $\bdy$ which itself is a $d$-dimensional Lorentzian manifold with a fixed metric.  Since we will want to consider points on the boundary as part of the bulk spacetime, we will also define the closure of $\bulk$, denoted $\cbulk = \bulk \cup \partial \bulk$, to be the spacetime including its boundary.
For definiteness we will consider globally static  boundary geometries ${\cal B}$, which admit a well defined foliation by fixed-time Cauchy slices; $\Sigma_{\cal B}$ will denote a typical spacelike leaf of such a foliation. Let us further pick the region ${\cal A}$  of interest to be a closed simply-connected sub-region on $\Sigma_{\cal B}$.
The boundary of ${\cal A}$, which clearly is likewise contained in the leaf $\Sigma_{\cal B}$, is denoted\footnote{
The surface $\partial {\cal A}$ is also sometimes referred to as the entangling surface.
}  as $\partial {\cal A}$.
We are interested in the following question: given field theory operators localized in ${\cal A}$, how much of the bulk spacetime $\bulk$ is accessible holographically to this region?

\begin{table}[tdp]
\begin{center}
\begin{tabular}{|c|c|c|c|c|}
\hline \hline
Notation & Meaning \& & dimen- &  open  / &  bulk  / \\
& Definition & sionality & closed & boundary \\
\hline\hline
${\cal M}$  
	& asymptotically locally AdS spacetime 
	& $d+1$ 
	& open 
	& bulk 
  \\  
$\cbulk$  
	& bulk spacetime including its boundary
	& $d+1$ 
	& closed 
	& bulk$+$bdy 
  \\  \hline
& conformal boundary of ${\cal M}$
	&&& \\
		$\partial {\cal M}$ or ${\cal B}$ 
		&  \& background for QFT& $d$ 
		& open 
		& boundary \\
	 &  foliated by constant time slices&  &&  
  \\ \hline
$\Sigma_{\cal B}$  
	& spacelike Cauchy slice on ${\cal B}$ 
	& $d-1$ 
	&  
	& boundary \\
${\cal A}$  
	& spacelike region contained in $\Sigma_{\cal B}$ 
	& $d-1$ 
	& closed
	& boundary  \\
$\partial {\cal A}$ 
	&  boundary of ${\cal A}$ (aka entangling surface)  
	& $d-2$ 
	& 
	& boundary 
  \\  \hline
$J^\pm(p)$
	& bulk domain of influence (including $p$)
	& $d+1$
	& closed$^\ast$
	& bulk$+$bdy
	\\ 
	& $\set{q \in \cbulk \st  \exists \ \curt{p \to q}^{\pm} \in \cbulk }$ 
	&&&
  \\  
$\domiy^\pm(p)$
	& boundary domain of influence (including $p$)
	& $d$
	& closed
	& boundary
	\\ 
	& $\set{q \in {\cal B} \st  \exists \ \curt{p \to q}^{\pm} \in {\cal B} }$ 
	&&&
  \\  
$\domdy^\pm[{\cal S}]$
	& boundary domain of dependence
	& $d$
	& closed
	& boundary
	\\ 
	& $\set{q \in {\cal B} \st
	  \forall \ \curt{q}^{\mp} \in {\cal B} , \
	  \curt{q}^{\mp} \cap {\cal S}  \ne   \emptyset  }$ 
	&&&
  \\  \hline
$\caustic^\pm$
	& ``caustics" which determine $\domdy^\pm[{\cal A}]$
	& $d-2$
	& closed
	& boundary
	\\ 
	& $\set{q \in \domdy^{\pm}[{\cal A}] \st (\domiy^\mp(q) \cap \domdy^{\pm}[{\cal A}]) \backslash q = \emptyset}$
	&&&
  \\  
$\domd$  
	& causal development of ${\cal A}$ within ${\cal B}$ 
	& $d$ 
	& closed
	& boundary  
	\\
	& $\domdy^+[{\cal A}] \cup \domdy^-[{\cal A}]
		=\domiy^-[\caustic^+] \cap  \domiy^+[\caustic^-] $
	&&&
  \\  
$\cwedge$ 
	& bulk causal wedge of  $\domd$ 
	& $d+1$ 
	& closed$^\ast$
	& bulk$+$bdy 
	\\ 
	& $J^-[\domd] \cap J^+[\domd]
		=J^-[\caustic^+] \cap  J^+[\caustic^-] $ 
	&&&
  \\  \hline
$\bcwedge$ &  boundary of the causal wedge in bulk & $d$ & &bulk\\
	& $\partial \cwedge \backslash \bdy$ 
	&&& \\
$\bcwedgef$ & future null boundary of the causal wedge & $d$ & &bulk\\
$\bcwedgep$ & past null boundary of the causal wedge & $d$ & &bulk\\
\hline
$\extr{\cal A}$ & extremal surface in bulk anchored on $\partial {\cal A}$ & $ d-1$ & & bulk \\
$\csf{\cal A}$ & causal information surface& $ d-1$ & & bulk \\
	& $\bcwedgef \cap \bcwedgep$ &&&\\
$\csfz{\cal A}$ & minimal-area surface on $\bcwedge$ & $ d-1$ & & bulk \\
\hline
\end{tabular}
\end{center}
\caption{Definition of the various regions and surfaces relevant to our construction.}
\label{t:defns}
\end{table}

The basic ingredients in our construction are causally defined domains, in particular domain of dependence and domain of influence.  In order to define these in a self-contained manner, 
we will first introduce the concept of causal curves; for a more detailed treatment, see e.g.\ \cite{Wald:1984ai}. A {\it causal} curve $\gamma$ is by definition a nowhere-spacelike (i.e., locally timelike or null) connected curve,  either in the (extended) bulk, $\gamma: {\mathbb R} \to \cbulk$, or confined to the boundary, $\gamma:{\mathbb R} \to \bdy$. 
 We take $\gamma$ (written without any subscript) to be maximally extended, in the sense that it can't just end inside the spacetime.  
 One step in our construction will also require us to distinguish future-   and past- directed causal curves.  To that end, we define $\curf p$ to be the future-directed causal curve starting at $p$,  and similarly  $\curp p$ the past-directed causal curve starting at $p$
(which is also a future-directed causal curve ending at $p$).
We will further  denote a future-directed causal curve  from a point $p$  to a point $q$ (which need not be distinct) in the spacetime by $\curf {p \to q} = \curp {q \to p}$.   Existence of such a curve guarantees that $q$ is in the future of $p$ (or equivalently $p$ is in the past of $q$).

To make this notion more precise, 
we will denote the bulk future {\it domain of influence} of a point $p$ by the standard notation $J^+(p)$ (and similarly denote past domain of influence of a point $p$ by $J^-(p)$).
These characterize the bulk regions which can  be influenced by  (or influence) the event $p$, respectively.  Technically, they consist of all points which lie in causal future (past) of $p$, i.e.,
\begin{equation}
J^+(p) =  \set{q \in \cbulk \st \exists \ \curf {p \to q} \subseteq \cbulk}
\ands
J^-(p) =  \set{q \in \cbulk \st \exists \ \curp {p \to q} \subseteq \cbulk}
\label{bulkdominfl}
\end{equation}	
Since the point $p$ is included in $J^\pm(p)$, these are typically\footnote{
A possible caveat is that under certain exotic (and rather unphysical) situations, $J^{\pm}(p)$ need not be closed, but rather may only be clopen.  This could happen for example in a spacetime with some point on the light cone of $p$ excised; however such situations are not physically relevant to our considerations, and we will therefore ignore them.
}
closed sets.
We can extend this notion to pertain to a region rather than a single point: a domain of influence of a given region ${\cal S}$ is simply the union of the domains of influence for all points in that region,
$J^+[{\cal S}] = \cup_{p\in {\cal S}} J^+(p)$, etc..  In particular, we can consider $J^\pm[{\cal A}]$ 
(where ${\cal A}$ is taken as a set in the extended bulk spacetime $\cbulk$), as illustrated in \fig{f:cwedge}.

If we restrict attention to the boundary spacetime ${\cal B}$ only, with $p \in {\cal B}$, causal curves on the boundary determine the {\it boundary domain of influence} of $p$, which we'll denote by $\domiy^\pm(p)$, with the obvious extension to boundary regions, such as $\domiy^\pm[{\cal A}]$.   In particular, $\domiy^\pm(p)$ is defined as in \req{bulkdominfl} with $\cbulk$ replaced by ${\cal B}$.
Although points $p \in \domiy^+[{\cal A}]$ can be influenced by ${\cal A}$, they are not determined by ${\cal A}$.  The latter set is more restricted, and corresponds to the boundary future domain of dependence $\domdy^{+}[{\cal A}]$, defined as the set of boundary points from which any past-directed causal curve on the boundary necessarily intersects ${\cal A}$, i.e.\
\begin{equation}
\domdy^{+}[{\cal A}]
= \set{q \in {\cal B} \st \forall \ \curp q \subseteq {\cal B},  \set{\curp q \cap {\cal A}}
\ne \emptyset } \ .
\label{}
\end{equation}	
Similar construction applies to the boundary past  domain of dependence $\domdy^{-}[{\cal A}]$.
The full boundary domain of dependence of ${\cal A}$ is then defined as
\begin{equation}
\domd
	=\domdy^{+}[{\cal A}] \cup \domdy^{-}[{\cal A}] \ .
\label{}
\end{equation}	
Specification of initial conditions in ${\cal A}$ fully determines the physics in the entire $\domd$.

We can now proceed to define the
 bulk causal wedge $\cwedge$ as the set of bulk points which lie in both future and past domains of influence of $\domd$, where $\domd  \subset {\cal B} \in \cbulk$,
\begin{equation}
\cwedge = J^-[\domd] \cap  J^+[\domd] \ .
\label{}
\end{equation}	
In fact the argument  of  $J^\pm$ can be reduced significantly: instead of the full $\domd$, we only need to consider the future-/past-most parts of $\domd$ whose past/future boundary domains of influence contain the entire $\domd$.
We'll denote these as $\caustic^+$ and $\caustic^-$, respectively, 
so that we can write
\begin{equation}
\domd = \domiy^-[\caustic^+] \cap  \domiy^+[\caustic^-] 
\thuss
\cwedge =  J^-[\caustic^+] \cap  J^+[\caustic^-] \ .
\label{}
\end{equation}	
Although we can take the left equality as the defining relation for $\caustic^\pm$, we can  also define $\caustic^\pm$ more explicitly by
\begin{equation}
\caustic^+ = \set{p \in \domd \st (\domiy^+(p) \cap \domd) \backslash p = \emptyset}
\ands
\caustic^- = \set{p \in \domd \st (\domiy^-(p) \cap \domd) \backslash p = \emptyset}
\label{}
\end{equation}	
Technically, $\caustic^{\pm}$ correspond to caustics\footnote{
Here we define {\it caustic} as set of intersections of null generators. Sometimes these are also referred to as ``crossover'', while ``caustic" is reserved for crossover of neighboring generators only; however, we will not need to adopt such refinements here.
} of the ingoing null geodesics emanating normal to the entangling surface $\entsurf$, and their construction and properties are discussed in greater detail in \sec{s:caustics}.
In general, they form (possibly branched) co-dimension 1 surfaces on the boundary of $\domd$ (so they are $(d-2)$-dimensional), but in special cases they can degenerate to lower-dimensional surfaces.  In the most symmetric case of  ${\cal A}$ being the round ball, corresponding caustics degenerate to a single point, which we denote by $\caustic^+ = \ddtipf$ and $\caustic^- = \ddtipp$.
In this special case, the causal wedge $\cwedge$ may be thought of as generated by Rindler horizons of bulk observers starting at $\ddtipp$ or ending at $\ddtipf$, so that
$\cwedge = I^-[\ddtipf] \cap  I^+[\ddtipp]$. 
This definition has a number of conceptual and technical advantages (since we are dealing with Rindler horizons), but we stress that except for $d=2$, this describes a highly-non-generic case.

Rindler horizons are by definition null surfaces; but it is true more generally that the boundary of a causal set (such as $J^\pm$ or $D^\pm$) is a null surface and is in fact generated by null geodesics (except possibly at a set of measure zero corresponding to the caustics of these generators).  For our bulk causal wedge $\cwedge$, it will similarly be true that its boundary in the bulk (i.e.\ with the AdS boundary itself removed), which we'll denote by $\bcwedge$, is composed of two null surfaces,
$\bcwedgef$ and $\bcwedgep$, which intersect along a spacelike surface $\csf{\cal A}$,
\begin{equation}
\bcwedge
	 = \bcwedgef \cup \bcwedgep 
\ands
\csf{\cal A}
	 = \bcwedgef \cap \bcwedgep \ .
\end{equation}
Hence  the surface $\csf{\cal A}$ is a spacetime co-dimension two bulk surface, which is by construction anchored on the entangling surface $\partial {\cal A}$ of the selected region ${\cal A}$, i.e.\ $\partial(\csf{\cal A})  = \partial {\cal A} $.

One can also consider the set of all spacelike surfaces lying in $\bcwedge$ and pinned at $\partial {\cal A}$. From these spacelike surfaces, we take the one with the minimal area. We denote this minimal
surface by $\csfz{\cal A}$.  So we have 
\begin{equation}
\csfz{\cal A}\subseteq \bcwedge  \subseteq \bulk \ \  , \qquad  \partial(\csfz{\cal A})  = \partial {\cal A} \ .
\label{}
\end{equation}	
We claim that in fact the two definitions given above coincide $\csfz{\cal A} = \csf{\cal A}$. 
This can be easily seen as follows:
Consider any surface $\Upsilon_{\cal A} \subseteq \bcwedge$.  We can obtain this surface from  $\csfz{\cal A}$ by flowing a certain distance $\lambda$ along the null generators.  Let us for the moment assume that  $\Upsilon_{\cal A}$ lies along the future-directed null generators of $\bcwedgef$; then we can perform a constant rescaling of the affine parameter of each null generator individually, such that $\Upsilon_{\cal A}$ lies at constant affine parameter $\lambda_0$ along the future-directed null generators of $\bcwedge$.
Now, we know that the expansion of the null generators of $\bcwedge$ cannot be negative towards the boundary (otherwise the generators would caustic, contradicting the fact that they reach the boundary along the null surface $\bcwedge$).  This means that the area of constant $\lambda$ slices of $\bcwedge$ must be monotonically increasing function of $\lambda$; in particular,
${\rm Area}(\Upsilon_{\cal A}) \ge {\rm Area}(\csfz{\cal A}) $.  Same argument would apply for $\Upsilon_{\cal A}$ lying on $\bcwedgep$, as the past-directed null generators again expand towards the boundary.
If $\Upsilon_{\cal A}$ lies partly on $\bcwedgef$ and partly on $\bcwedgep$,  then we can separate $\csfz{\cal A}$ into domains, separated by $\Upsilon_{\cal A} \cap \csfz{\cal A}$, and run the argument for each domain separately.  Hence in all cases, any  surface $\Upsilon_{\cal A}$ cannot have smaller area than $ \csfz{\cal A}$, which means that $ \csfz{\cal A}$ is the minimal surface on $\bcwedge$, i.e.\ $\csfz{\cal A} = \csf{\cal A}$. 

We note in passing that the construction does not depend on a choice of coordinates, but only on physically meaningful quantities: causal relations in the spacetime.  This ensures that we can apply the same construction even for time dependent bulk geometries. More importantly this works in any theory of dynamical gravity satisfying sensible energy conditions; for instance higher derivative theories of gravity that have attracted some interest recently will admit exactly the same conditions since the construction is predicated on causal relations alone.

Before moving on, let us note that the causal construction of the surface $\csf{\cal A}$  described above 
was originally considered in \cite{Hubeny:2007xt} in the context of holographic entanglement entropy. This surface was called ${\cal Z}$ in that work which also erroneously declared it to be the maximal area surface. As we shall see later, the area of spacelike surfaces measured along $\bcwedge$ at fixed affine parameter of the null generators  increases as we move away from $\csf{\cal A}$ towards the boundary, implying that there are indeed surfaces with larger area than that of $\csf{\cal A}$ on $\bcwedge$. 

\subsection{Defining causal holographic information}
\label{s:chi}

Having constructed the surface $\csf{\cal A}$ we define the following measure of holographic information associated with the region ${\cal A}$:
\begin{equation}
\chi_{\cal A} = \frac{\text{Area}(\csf{\cal A})}{4\,G_N} 
\label{}
\end{equation}	
We want to claim that $\chi_{\cal A}$ provides a lower  bound on the information regarding the bulk that the region ${\cal A}$ has. The causal nature of the construction which guides our intuition makes this plausible, but it would be useful to have a first-principles proof of this statement. 

A-priori  it seems plausible that $\chi_{\cal A}$ is some functional of the reduced density matrix $\rho_{\cal A}$. After all, the knowledge of $\rho_{\cal A}$ is sufficient to compute field theory observables in $\domd$.  However,  we will see later that generically $\chi_{\cal A} \neq \chi_{{\cal A}^c}$, where ${\cal A}^c$ is the complementary region to ${\cal A}$. This happens even for pure states of the entire system on  $\Sigma_{\cal B}$, for which the spectrum of eigenvalues of $\rho_{\cal A}$ and $\rho_{{\cal A}^c}$ are identical. These observations might lead one to conclude that $\chi_{\cal A}$ cannot  be determined by knowledge of $\rho_{\cal A}$ alone,\footnote{ We thank Hong Liu for alerting us to this possibility.}  which manifestly contradicts the original assertion that $\rho_{\cal A}$ determines observables in the $\domd$ and thence in $\cwedge$. However, we are overlooking the fact the entanglement spectrum (i.e., set of eigenvalues) of $\rho_{\cal A}$ only determines the reduced density matrix  up to unitary transformations. It is therefore possible that we can deduce $\chi_{\cal A}$ from $\rho_{\cal A}$, though the precise nature of this dependence and an intrinsic field theoretic definition of $\chi_{\cal A}$  is an interesting open question which we leave for the future.

\subsection{Lightning review of holographic entanglement entropy}
\label{s:entreview}

 The computation of holographic entanglement entropy also requires specification of some region ${\cal A}$ on a spatial slice and is defined in terms of the von Neumann entropy of the reduced density matrix $\rho_{\cal A}$, see \req{eedef}. If the entire state of the quantum field theory is static, then the reduced density matrix is time-independent; relatedly the bulk holographic dual spacetime is also static. For these situations, we can compute the entanglement entropy using the minimal surface prescription of \cite{Ryu:2006ef}. However, we can be more general, and consider states that have non-trivial time dependence, so that the density matrix is not time-independent. In such cases the foliation by spacelike surfaces $\Sigma_{\cal B}$ which exists on the rigid field theory background ${\cal B}$ does not necessarily extend to the bulk in a unique way. As a result, it does not suffice to consider minimal surfaces, but rather as explained in \cite{Hubeny:2007xt} one looks at extremal surfaces $\extr{\cal A}$ which is an extremum of the area functional (these surfaces were denoted as ${\cal W}$ in \cite{Hubeny:2007xt}). This extremal surface  is anchored on the boundary $\partial {\cal A}$ of the region ${\cal A}$ and the entanglement entropy is given as\footnote{
 At this point we are restricting attention to two derivative theories of gravity in the bulk. If higher derivative curvature terms are present in the bulk then we need to consider a suitably generalized functional as has been discussed in \cite{deBoer:2011wk,Hung:2011xb}. Similar considerations should also apply to $\chi_{\cal A}$.}
\begin{equation}
S_{\cal A} = \frac{\text{Area}(\extr{\cal A})}{4\,G_N} 
\label{}
\end{equation}	
%

\section{Holographic information}
\label{s:connections}

We now have at hand two different constructions in the bulk associated with the given boundary region ${\cal A}$. Both the causal wedge $\cwedge$ (and therefore its associated  co-dimension  two causal information surface $\csf{\cal A}$) as well as the extremal surface $\extr{\cal A}$ are constructed covariantly, without any preferred choice of bulk foliation etc.. We now proceed to explain some of the features of the causal construction, focussing on its relation to the extremal surface in particular. 

\subsection{Causal information  versus entanglement entropy }
\label{s:}
Let us first address the following question: why is it that $\csf{\cal A}$ is not the correct candidate to compute entanglement entropy of the dual field theory? The basic argument for this was already presented in Appendix A of \cite{Hubeny:2007xt} which we now review.

First of all, let us note that the causal construction depends on the causal structure of the bulk spacetime $\bulk$ and thus is oblivious to the conformal structure of the bulk metric. This in particular means that the causal construction in pure \AdS{d+1} geometries in the Poincar\'e patch with the metric
\begin{equation}
ds^2 = \frac{-dt^2 + d{\bf x}^2 + dz^2}{z^2} 
\label{}
\end{equation}	
is the same as the construction of causal wedges in flat spacetime ${\mathbb R}^{d,1}$, restricted to the half-space $z > 0$. For instance, if we take ${\cal A}$ to be a strip-like region of width $w$:
\begin{equation}
 {\cal A} = \{ {\bf x} : |x_1| \leq \frac{w}{2}, \; x_i \in {\mathbb R},  \ \text{for} \ i = 2, \ldots, d-1 \}  
\label{stripdef}
\end{equation}	
one can check that $\csf{\cal A}$ is simply given by the half-cylinder \cite{Hubeny:2007xt}
\begin{equation}
 \csf{\cal A} = \{ ({\bf x}, z) : z^2 + x_1^2 = \left(\frac{w}{2}\right)^2 , \;z >0 , \;x_i \in {\mathbb R},  \ \text{for} \ i = 2, \ldots, d-1 \}  
\label{csfstrip}
\end{equation}	
On the other hand, the minimal surface $\extr{\cal A}$ is a more complicated surface constructed in \cite{Ryu:2006bv}, defined by:
\begin{equation}
 \frac{dz}{dx_1} = \frac{\sqrt{(z_*)^{2(d-1)} - z^{2(d-1)}}}{z^{d-1}} \ , \qquad z_* = \frac{\Gamma\left(\frac{1}{2(d-1)}\right)}{\sqrt{\pi}\, \Gamma\left(\frac{d}{2(d-1)}\right)} \; \frac{w}{2}
\label{extrstrip}
\end{equation}	
where $z_*$ corresponds to the depth (maximal radial coordinate) to which this surfaces reaches.  Written more explicitly (cf.\ eg.\ \cite{Hubeny:2012ry}), one can express $x(z)$ explicitly as
\begin{equation}
\pm x(z) =  \frac{z^{d}}{d \, {z_*}^{d-1}} \ {}_2 F_1 
\left[\frac{1}{2}, \frac{d}{2(d-1)},  \frac{3d-2}{2d-2}, \frac{z^{2(d-1)}}{z_*^{2(d-1)}} \right]
- z_* \,  \frac{\sqrt{\pi} \, 
\Gamma \left[ \frac{3d-2}{2d-2} \right]}{d \,  \Gamma \left[ \frac{2d-1}{2d-2} \right]} \ .
\label{}
\end{equation}	
The easiest way to confirm that the two surfaces, $\csf{\cal A}$ and $\extr{\cal A}$, cannot coincide when $d>2$ is to note that in particular they reach to different depth in the bulk.  Whereas the causal information surface  $\csf{\cal A}$ always reaches only to 
$z_*^{(\Xi)} = w/2$, the reach of the extremal surface depends on dimensionality of the surface, and is given by \req{extrstrip}, where it can be easily confirmed that the coefficient of $w/2$ is always greater than 1 for any $d>2$ and increases with $d$; in other words, the extremal surface reaches deeper than the causal information surface.  (In $d=2$ the two expressions coincide since both surfaces are one-dimensional and correspond to a spacelike geodesic anchored at the endpoints of ${\cal A}$.)

Likewise, it is easy to check that $\chi_{{\cal A}}$ and $S_{{\cal A}}$ differ.
Evaluating the areas for two surfaces we find:
\begin{eqnarray}
\chi_{\cal A}
&=&  2\,c_\text{eff} \; \left(\frac{2\,L}{w}\right)^{d-2}\;  \sqrt{1-\frac{4\,\varepsilon^2}{w^2}} \  \ _2F_1\left(\frac{1}{2}, \frac{d}{2}, \frac{3}{2}, 1-\frac{4\,\varepsilon^2}{w^2}\right) 
\nonumber \\
&=&  2\, c_\text{eff}\; L^{d-2}  \left[\frac{1}{(d-2)} \; \frac{1}{\varepsilon^{d-2}} +  \frac{f_{d-4}}{w^2\,\varepsilon^{d-4}} +\cdots +
\begin{cases}
& f_0   \qquad \qquad \quad \;\; d \ \text{odd} \\
& \frac{f_a}{w^{d-2}} \, \log\left(\frac{w}{\varepsilon}\right)  \quad d \ \text{even} 
\end{cases}
\right] 
\label{chistrip}
\end{eqnarray}	
which turns out to contain logarithmically divergent terms for even $d$.  Here we define $c_\text{eff} = \frac{ \ell^{d-1}}{4\, G_N} $, with $\ell$ being the \AdS{d+1} scale, as a measure the effective central charge of the dual field theory.\footnote{
The central charge is usually defined more conventionally as $c = \frac{1}{4\pi}\, c_\text{eff}$ taking into account the conventional normalization of the Einstein-Hilbert action. For ${\cal N}=4$ SYM we have for instance $c_\text{eff} = \frac{N^2}{2\pi}$.}
$L$ is an infra-red regulator we introduced to compute the area along the non-compact directions of the strip by restricting $|x_i| \leq L$  for $ i \neq 1$, while $\varepsilon$ is a UV cut-off introduced to extract the divergent terms in the area. The coefficients $f_i$ are easily obtained by expanding out the hypergeometric function. On the other hand we find that the minimal surface area leads to 
\begin{equation}
S_{\cal A}= \frac{c_\text{eff}}{d-2} \,\left[2\, \left(\frac{L}{\varepsilon}\right)^{d-2} - \, \left(\frac{2\,\sqrt{\pi}\;\Gamma\left(\frac{d}{2(d-1)}\right)}{\Gamma\left(\frac{1}{2(d-1)}\right)}\right)^{d-1} \, \left(\frac{L}{w}\right)^{d-2}\right]
\label{eestrip}
\end{equation}	
which only contains a finite piece apart from the divergent part proportional to the area of the strip.
The main observation made in \cite{Hubeny:2007xt} was that the result \req{eestrip} was similar in structure to the result obtained in free gauge theories, while \req{chistrip} contains extra divergent pieces.\footnote{
The difference in the divergence structure of the area of the surfaces $\csf{\cal A}$ and $\extr{\cal A}$ can be traced to the fact that they approach the boundary differently. While both surfaces have to hit the boundary normally since the spacetime is asymptotically \AdS{}, the subleading pieces differ due to the different geometric constructions, producing a relative bending between them. This raises the question whether one should compare the two answers with a regulator that accounts for the relative bending as opposed to the rigid UV cut-off which we have employed above. We thank Juan Maldacena for discussions on this issue.}
For instance, in the case $d=4$, we have \MR{footnote}
\begin{equation}
 S_{\cal A}= c_\text{eff}\,L^2\, \left(\frac{1}{\varepsilon^2} - \frac{0.32}{w^2}\right) \ , \qquad  
 \chi_{\cal A} = c_\text{eff}\, L^2\,\left(\frac{1}{\varepsilon^2} - \frac{2}{w^2} + \frac{4}{w^2}\, \log\left(\frac{w}{\varepsilon}\right)\right) \,
\label{entchistrip}
\end{equation}	
while free ${\cal N}= 4$ SYM gives \cite{Ryu:2006bv}
\begin{equation}
S_{\cal A}^{\text{free}} = c_\text{eff}\,L^2\, \left(\frac{1}{\varepsilon^2} - \frac{0.49}{w^2}\right) \ .
\label{}
\end{equation}	
The absence of the logarithmically divergent term in the free result and the area of the minimal surface suggests that the strong coupling answer for the holographic entanglement entropy is given by the minimal surface prescription, and not the causal construction.

This argument however does not take into account potential strong coupling effects that can arise and perhaps induce subleading divergent terms in entanglement entropy. Thus far there is no explicit computation in a strongly coupled theory that can independently assert the absence of such terms. 

However, we can formulate a more robust argument for why the causal construction cannot  give the entanglement entropy in general. There are two basic observations we will employ to establish this result.

\paragraph{1. Causal information of region ${\cal A}$ and its complement:} Let us consider a QFT in a pure state $\ket{\Psi}$. We pick a time-slice $\Sigma_{\cal B}$ and on this slice and mark off the region ${\cal A}$. We then have the complement of the region ${\cal A}^c  = \Sigma_{\cal B}\backslash {\cal A}$, whose reduced density matrix could be computed by integrating out the degrees of freedom in $ {\cal A}$. Since the total state is pure, it follows from the definition of the entanglement entropy that $S_{\cal A}= S_{{\cal A}^c}$. 
This is easily shown to be true for the extremal surface construction, since there is a single extremal surface $\extr{{\cal A}}=\extr{{\cal A}}^c$ in the bulk that lies anchored on the boundary $\partial {\cal A}$ which is the common boundary of both  ${\cal A}$ and ${\cal A}^c$.\footnote{
In making this argument, we use the fact that pure states in the field theory correspond to horizon-free geometries in the bulk; thus the homology constraint described in \cite{Headrick:2007km} plays no role in our discussion.} 

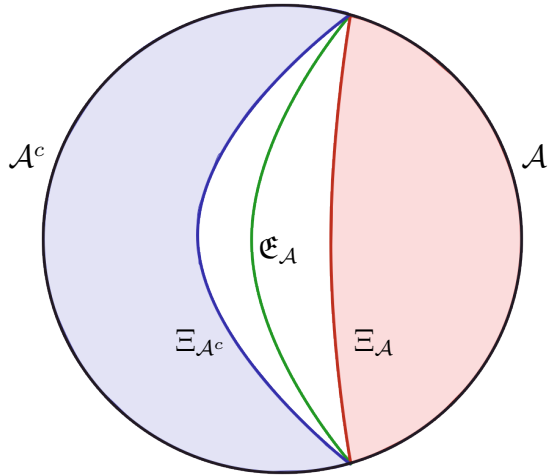
\begin{figure}
\begin{center}
\includegraphics[width=2.5in]{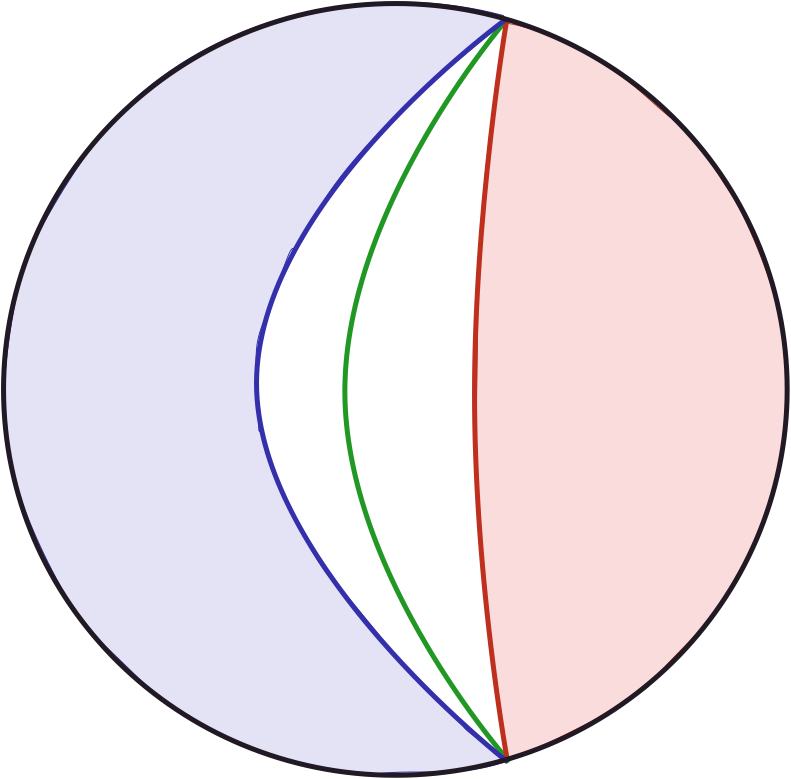}
\begin{picture}(1,1)(0,0)
\put(-190,120){\makebox(0,0){${\cal A}^c$}}
\put(0,120){\makebox(0,0){${\cal A}$}}
\put(-60,50){\makebox(0,0){$\csf{{\cal A}}$}}
\put(-125,50){\makebox(0,0){$\csf{{\cal A}^c}$}}
\put(-96,85){\makebox(0,0){$\extr{{\cal A}}$}}
\end{picture}
\caption{Sketch to illustrate the fact the causal information surfaces $\csf{\cal A}$ and $\csf{{\cal A}^c}$ for a region ${\cal A}$  and its complement  ${\cal A}^c$ have to lie closer to the respective boundary regions than the common extremal surface $\extr{\cal A} = \extr{{\cal A}^c}$.
}
\label{f:xicomplement}
\end{center}
\end{figure}

However, for the causal construction there is an asymmetry generically between the causal wedges of the regions ${\cal A}$ and ${\cal A}^c$.\footnote{
This argument was developed together with Mark van Raamsdonk.} 
The basic point is quite simple and the main idea is sketched in \fig{f:xicomplement},  set in the more natural context of global AdS.  Consider e.g.\ a static asymptotically global AdS geometry with a gravitational potential well.  By the Gao-Wald theorem \cite{Gao:2000ga}, within a fixed time set by the size of $\domd$, the null geodesics which define the causal wedge cannot reach as far from the AdS boundary as they could in the pure AdS spacetime.  But in pure global AdS, the causal information surfaces for a {\it circular} region ${\cal A}$ and its complement would coincide.\footnote{
The reason is apparent from \fig{f:cwedge3d}(a), where the null boundaries of the causal wedge for ${\cal A}$ corresponding to half the circle are shown.  These are Rindler horizons, and due to the large symmetry Rindler horizons from any other point would look the same.  In particular, to construct causal wedge for any other circular region (i.e.\ shorter interval in  \fig{f:cwedge3d}(a)), we can  simply  time-translate one of the null planes with respect to the other.  But in pure AdS, the same null plane acts both as the past boundary of ${\cal A}$'s causal wedge and as the future boundary of ${\cal A}^c$'s causal wedge, since null geodesics through AdS all reconverge at the same antipodal null-translated point.  Since  the two null planes (future and past boundaries of either region's causal wedge) always intersect on a single surface; this  surface is simultaneously $\csf{{\cal A}}$ and $\csf{{\cal A}}^c$.  
}  
Hence for any physical deformation of AdS, the causal information surfaces would shift, $\csf{{\cal A}}$ towards the boundary where ${\cal A}$ is located, and $\csf{{\cal A}}^c$ towards the boundary where ${\cal A}^c$ is located, as indicated in \fig{f:xicomplement}.
Moreover, due to caustics in $\domd$ for any other shaped region in $d>2$, the corresponding causal information surfaces would likewise retreat towards the boundary, even for pure AdS, whenever ${\cal A}$ is not the round ball.  Thus, in general, $\csf{{\cal A}}$  and $\csf{{\cal A}}^c$ differ, so there is no reason for $\chi_{\cal A}$  and $ \chi_{{\cal A}^c}$ to be the same.
  
To see an explicit example, for simplicity in the context of flat boundary, let us again consider the strip discussed above; but in order to keep both ${\cal A}$ and its complement finitely extended in at least one direction, let the  $x_1$ direction be compactified, say $x_1 \sim x_1 +  R$.  This means we should consider the boundary theory on ${\mathbb R}^{d-2,1} \times {\bf S}^1$ and let $\ket{\Psi}$ be the corresponding vacuum state. Its bulk dual is then again Poincar\'e \AdS{d+1}, but now with  the $x_1$ direction compactified. 
Now consider the strip-like region \req{stripdef} as the region ${\cal A}$, so that the complement is simply the strip with a longer arc of length $R-a$. In this case, it is easy to see that the causal developments of the two regions are different: $ \domd \neq \domdc$. This per se is not sufficient to guarantee that the surface of interest $\csf{\cal A}$ and $\csf{{\cal A}^c}$ are different (as we saw above, for the example of pure global AdS). However, for these strip like regions, it is easy to show that the surfaces $\csf{\cal A}$ and $\csf{{\cal A}^c}$ are completely different. The former is given by a half-cylinder \eqref{csfstrip} with circumference  $a$ while the latter is  a half-cylinder with circumference $R-a$. Using \eqref{chistrip} we then clearly see that 
$\chi_{\cal A}\neq \chi_{{\cal A}^c}$. We have already commented on the implications of this observation in \sec{s:chi}.

\paragraph{2. $\chi$ is not sub-additive:} A further issue with $\chi_{\cal A}$ is that it does not satisfy an appropriate convexity property, called strong sub-additivity \cite{Lieb:1973cp}.  Usually this is written for two regions ${\cal A}_1$ and ${\cal A}_2$ and demands that 
\begin{eqnarray}
S_{{\cal A}_1} + S_{{\cal A}_2} &\geq& S_{{\cal A}_1 \cup {\cal A}_2} + S_{{\cal A}_1 \cap {\cal A}_2} 
 \nonumber \\
S_{{\cal A}_1} + S_{{\cal A}_2} &\geq& S_{{\cal A}_1 \backslash {\cal A}_2} + S_{{\cal A}_2 \backslash {\cal A}_1}  
 \label{subadd}
\end{eqnarray}	
Strong sub-additivity is  a property that is satisfied naturally by any von Neumann entropy of a density matrix, and in particular holds for the reduced density matrix.

Let us turn to the holographic constructions: for static states in the field theory, using the minimal surface prescription of \cite{Ryu:2006bv}, a very elegant proof of holographic entanglement entropy being sub-additive was given by \cite{Headrick:2007km}. This construction relies on the fact that the surfaces of interest in the static case are minimal surfaces and uses this to fact to derive the inequality \eqref{subadd} by a simple deformation argument. On the other hand, the  extremal surfaces relevant for  time-dependent states have not been amenable to a derivation of sub-additivity; while we expect that holographic entanglement entropy given by extremal surface area does satisfy \eqref{subadd}, such result has eluded proof to date.

However, it is possible to show that the  causal holographic information {\it cannot} satisfy \eqref{subadd}, by explicitly constructing counter-examples.  Focusing again on the simple case of strip-like regions, now overlapping, we can easily find the relevant ingredients.  Let ${\cal A}_1$ and ${\cal A}_2$ be strip-like regions \eqref{stripdef} for a field theory on ${\mathbb R}^{d-1,1}$, defined as
\begin{eqnarray}
 {\cal A}_1 &=& \{ {\bf x} : -\left(a_1 +\frac{x_0}{2} \right)\leq x_1 \leq \frac{x_0}{2}, \; x_i \in {\mathbb R},  \ \text{for} \, i \neq 1\} 
  \nonumber \\
  {\cal A}_2 &=& \{ {\bf x} : -\frac{x_0}{2} \leq x_1 \leq \left(a_2 + \frac{x_0}{2}\right), \; x_i \in {\mathbb R},  \ \text{for} \, i \neq 1\} \ ,
\label{}
\end{eqnarray}	
so that 
${\cal A}_1$ is a strip of width $x_0 + a_1$ and ${\cal A}_2$ is a strip of width $x_0 + a_2$, while 
${\cal A}_1 \cup {\cal A}_2 $ has width $x_0 + a_1 +a_2$ and ${\cal A}_1 \cap {\cal A}_2$ has width $x_0$ respectively. It is then easy to write down the expressions for the causal information for each of the regions in question and check that the sub-additivity property \eqref{subadd} is not satisfied. For instance in $d=4$, we would require that 
\begin{equation}
 F(a_1+x_0) + F(a_2 + x_0) - F(a_1 + a_2 + x_0) - F(x_0) > 0 \,, \qquad 
F(x) = \frac{1}{x^2}\, \log\left(\frac{x}{{\tilde \varepsilon}}\right)
\label{}
\end{equation}	
(where ${\tilde \varepsilon}$ is a rescaled cutoff which we can take to be some small number).
This inequality is clearly violated (choose for example for $x_0 = a_1 = a_2$). 
This adds further evidence that the causal information is not a von Neumann entropy for a general density matrix. This per se does not imply that $\chi_{\cal A}$ cannot be thought of as a measure of information: for e.g., the Renyi entropies \eqref{eedef} which capture the moments of the reduced density matrix, likewise fail to be sub-additive.

\subsection{Concordances: when $\csf{\cal A}$ and $\extr{\cal A}$ coincide }
\label{s:}

Thus far we have established that the causal holographic information $\chi$, whilst being very naturally related to the spatial region of interest in the boundary, lacks properties to make it a well behaved notion of (von Neumann) entropy. So the natural question is why do we bother with this concept? Clearly, entanglement entropy is a more natural object which we can associate directly to the reduced density matrix. However, as we have already suggested in \sec{s:intro}, there is a sense in which we expect that the bulk causal wedge $\cwedge$ is the minimal bulk region which should be reconstructable from the data on ${\cal A}$. It seems then natural from the bulk gravitational viewpoint to consider $\chi_{\cal A}$ as encoding  the amount of information (of a certain type) contained in the region ${\cal A}$.  

Before we get to our justification for using $\chi$ as a measure of information about the bulk spacetime, let us pause to explain some situations  where there is an explicit agreement between $\chi_{\cal A}$ and $S_{\cal A}$. These examples provide some interesting insight into the nature of $\chi$. 

\subsubsection{ $1+1$ dimensional CFTs on ${\bf S}^1$} 
\label{s:threed}

The first class of examples we have at our disposal are $1+1$ dimensional CFTs. We tabulate the three density matrices of interest and their corresponding holographic dual geometries in Table \ref{t:adsthree}. 
\begin{table}[h!]
\begin{center}
\begin{tabular}{|c|l |l|}
 \hline
Case & CFT density matrix &Holographic duals \\
\hline \hline
 (a).&CFT vacuum &   global \AdS{3} \eqref{ads3} \\
\hline
 (b).&Thermal density matrix & static BTZ geometry \eqref{btz} \\
\hline
 (c).& Grand canonical density matrix & rotating BTZ spacetime  \eqref{rbtz} \\
\hline 
\end{tabular}
\caption{Density matrices of interest in the $1+1$ dimensional CFT.}
\label{t:adsthree}
\end{center}
\end{table}
In the field theory we denote the Hamiltonian generator by $H$ and the angular momentum along ${\bf S}^1$ by $J$. 
The geometries of relevance for the cases considered in Table \ref{t:adsthree} are given to be:
\begin{align}
&\ket{\Psi} =\; \ket{0}:  
& ds^2  & = -\left(r^2+1\right)\, dt^2 + \frac{dr^2}{r^2+1} + r^2\, d\varphi^2
\label{ads3} \\
&\rho = e^{-\beta\, H} :
&ds^2  &= -\left(r^2 -r_+^2\right) dt^2 + \frac{dr^2}{r^2 - r_+^2} + r^2\, d\varphi^2 
\label{btz}\\
&\rho = e^{-\beta\, (H -\Omega\, J)}:
&ds^2  &= -\frac{(r^2 -r_+^2)(r^2-r_-^2)}{r^2}\, dt^2 + \frac{r^2\, dr^2}{(r^2 - r_+^2)(r^2-r_-^2)} + r^2\, \left(d\varphi + \frac{r_+\,r_-}{r^2}\, dt \right)^2 
 \label{rbtz}
 \end{align}
with 
\begin{equation}
\beta = \frac{2\pi}{r_+}
\label{}
\end{equation}	
for the static BTZ geometry and 
\begin{equation}
\beta \, (1\pm \Omega) = \frac{2\pi}{r_+ \pm r_-} \equiv \beta_\pm
\label{}
\end{equation}	
for the rotating BTZ spacetime.
  
We take the CFT to live on a circle of radius unity and further take the region ${\cal A}$ to be an arc of length $2\,\varphi_0$:
\begin{equation}
{\cal A} = \{ (t,\varphi) \;| \;  t =0,\  \varphi \in (-\varphi_0, \varphi_0) \} \ .
\label{regads3}
\end{equation}	
For such states with the choice of the region detailed above, we actually have a microscopic understanding of entanglement entropy. Using the replica trick, the entanglement entropy in cases (a) and (b) was originally obtained in \cite{Calabrese:2004eu}, while this argument was subsequently generalized to case (c) in \cite{Hubeny:2007xt}.  In all cases the computation relies on mapping the computation of ${\rm Tr}(\rho_{\cal A}^n)$ into the computation of the two point function of a twist operator. 

Let us quickly review the holographic computation of entanglement entropy in these cases. For the static situations in the vacuum or the thermal density matrix, we can use the minimal surface prescription of \cite{Ryu:2006bv}; in this low dimension the minimal surface is simply a geodesic. It is a simple matter to show from the geodesic equation that the bulk geodesic of interest is given by
\begin{align}
&\text{(a)}.& \extr{\cal A}:  && t & = 0\,, &r^2 (\varphi) &= \frac{\cos^2 \varphi_0 }{ \sin^2 \varphi_0 \, \cos^2 \varphi - \cos^2 \varphi_0 \, \sin^2 \varphi } 
\label{extrads3} 
\\
&\text{(b)}.& \extr{\cal A}: && t&=0\,, & r^2 (\varphi) &= r_+^2 \
{ \cosh^2 ( r_+ \, \varphi_0)  \over \sinh^2
  ( r_+ \, \varphi_0) \, \cosh^2  ( r_+ \, \varphi)
   - \cosh^2  (r_+ \, \varphi_0) \, \sinh^2  ( r_+ \, \varphi) }
 \label{extrbtz}
\end{align}	
For the stationary situation of the grand canonical density matrix one still looks for spacelike geodesics anchored at the end-points of the region ${\cal A}$  \eqref{regads3}; as described in \cite{Hubeny:2007xt} this can be obtained efficiently by passing to new coordinate where the metric \eqref{rbtz} is brought to Poincar\'e \AdS{3} form. To be specific, consider the coordinate transformation, 
\begin{align}
w_\pm &= \sqrt{\frac{r^2-r_+^2}{r^2-r_-^2}} \; e^{(r_+\pm\, r_-)\, \left(\varphi \pm t\right)} \equiv X \pm T\ 
\nonumber \\ 
z &=  \sqrt{\frac{r_+^2-r_-^2}{r^2-r_-^2}} \; e^{r_+\, \varphi + r_- \, t }
\label{wzdefs}
\end{align}	
which recasts \eqref{rbtz} into the form:
\begin{equation}
ds^2  =  \frac{dw_+\, dw_- + dz^2}{z^2} \ .
\label{pads3}
\end{equation}	
It is now easy to  obtain the extremal surface of interest in \eqref{rbtz} by boosting the Poincar\'e \AdS{3} geodesics.\footnote{
On the Poincar\'e disc  \eqref{pads3} spacelike geodesics on a constant  $T$ slice  are half-circles parametrized by their radius $h$:  $ (X - X_*)^2  + z^2 =  h^2$.} It turns out the desired spacelike geodesics live on a co-dimension one spacelike hypersurface of the bulk spacetime given by
\begin{align}
&\text{(c)}.& \qquad \gamma\, w_+ - \gamma^{-1} \, w_ - = \text{constant}  \,, \qquad 
\gamma^2 = \frac{\sinh(r_+ - r_-) \varphi_0}{\sinh(r_+ + r_-)\varphi_0} \qquad \qquad\qquad
\label{extrrbtz}
\end{align}	
in the Poincar\'e coordinates. The explicit expression for the geodesic is unilluminating and so we refrain from writing it down.

\begin{figure}
\begin{center}
\hspace{3mm}
\includegraphics[width=2.1in]{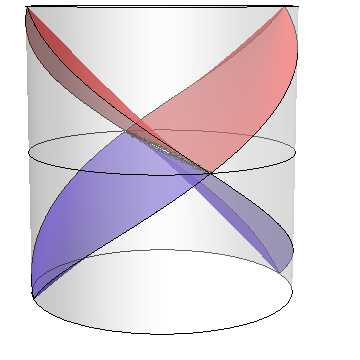}
\includegraphics[width=2.1in]{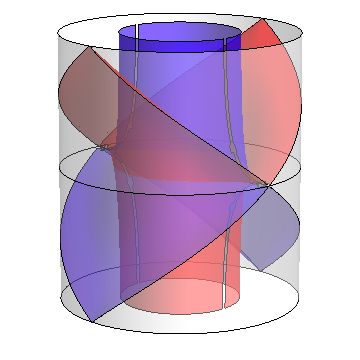} 
\hspace{3mm}
\includegraphics[width=2.1in]{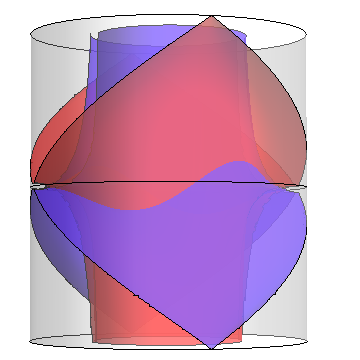}
\begin{picture}(1,1)(0,0)
\put(-410,-7){\makebox(0,0){\bf (a)}}
\put(-250,-7){\makebox(0,0){\bf (b)}}
\put(-85,-7){\makebox(0,0){\bf (c)}}
\end{picture}

\caption{
Illustration of the causal wedges $\cwedge$ in three dimensional asymptotically globally \AdS{3} spacetimes. The three figures correspond to the three geometries described in Table \ref{t:adsthree}.
For convenience we have chosen the region ${\cal A}$ to be a half of the boundary ${\bf S}^1$, i.e., $\varphi_0 = \pi$. At the intersection of the $\bcwedgef$ and $\bcwedgep$ lies the causal information surface $\csf{\cal A}$ which as we discuss in the text is the same as the extremal surface $\extr{\cal A}$ in these examples. Note that for the static spacetimes (a) and (b) which correspond to AdS{3} and the static BTZ geometry, the surfaces at a fixed time slice $t=0$ as shown, while for the stationary rotating BTZ geometry (c), this surface dips above and below the $t=0$ slice in the bulk. [Note that for ease of visualization, we have changed the viewpoint between the three plots.  Also, note that the `seams' are just numerical glitches.]}
\label{f:cwedge3d}
\end{center}
\end{figure}

Now we turn to the computation of the causal holographic information. In all the three situations of interest, the boundary geometry is a Lorentzian cylinder ${\mathbb R}\times {\bf S}^1$. For the region ${\cal A}$ in \eqref{regads3} the causal domain of dependence on the boundary is simply the diamond shaped region
\begin{equation}
\domd = \{(t,\varphi)  \;| \; t  \leq  |\varphi_0 - \varphi| \,, \varphi \in ( 0,\varphi_0) 
\;\; \cup \;\;  t \leq |\varphi_0 +\varphi| \,, \varphi \in (-\varphi_0, 0)  \}
\label{domads3}
\end{equation}	
The computation of the bulk causal wedge associated with \eqref{domads3} is straightforward; in fact we only need information about the Rindler horizons associated with this causal wedge for the rest of the computation. We use the symmetries of the problem to realize that we simply need to construct the past light cone from the top of $\domd$ i.e, from $(t=\varphi_0, \varphi=0)$  to obtain $\bcwedgef$ and the future light cone from the bottom of $\domd$, i.e., from 
 $(t=-\varphi_0, \varphi=0)$ to obtain $\bcwedgep$.  These can be done by explicitly writing bulk null geodesics emanating from these points. However, now we have to treat the three geometries in turn; the results of the construction outlined below are illustrated in \fig{f:cwedge3d} where we have explicitly plotted the causal wedges for the three cases in the AdS cylinder. 
 
\noindent
{\bf (a).} For global \AdS{3} parameterizing the geodesics by their conserved angular momentum $j$ we obtain:
\begin{eqnarray}
\label{tphgeodAdS}
t(r) &=&  \varphi_0  - {\pi \over 2}+  \tan^{-1} \sqrt{(1-j^2) \, r^2\, \ell^{-2} - j^2} \\
\varphi(r) &=&   \frac{\pi}{2} -  \tan^{-1}\frac{ \sqrt{(1-j^2) \, r^2 \,\ell^{-2}- j^2}}{ j }
\end{eqnarray}
Inverting and substituting to obtain both $r$ and $\varphi$ in terms of $t$ for any $j$, we get
\begin{equation}  \label{rphAdS}
\bcwedgef: \qquad r^2(t,j) = \frac{\cot^2 \left(\varphi_0-t\right)+ j^2}{1 -j^2} \ , \qquad
\varphi(t,j) = \frac{\pi}{2} - \tan^{-1} \left(\frac{\cot \left(\varphi_0-t\right)}{j} \right)
\end{equation}
Similar expression for $\bcwedgep$ can be written down by simply flipping the starting point of the geodesics to the point $(-\varphi_0,0)$.

Finally, to  obtain the surface $\csf{\cal A}$ we realize that all we need to do owing to the symmetries of the geometry is to look at the spacelike surface at $t=0$ on $\bcwedge$. Essentially one inverts the second expression to obtain $j$ in terms of $\varphi$: 
$j = \cot \varphi_0 \, \tan \varphi$, and substitutes back into $r$ to get $\csf{\cal A}$:
\begin{align} 
&\text{(a).}&  \csf{\cal A}: && t & = 0\,,
& r^2 (\varphi) =\frac{\cos^2 \varphi_0}{\sin^2 \varphi_0 \, \cos^2 \varphi - \cos^2 \varphi_0 \, \sin^2 \varphi}
 \label{csfads3}
\end{align}
which indeed agrees with the minimal surface \eqref{extrads3}.

\noindent
{\bf (b).} For the static BTZ geometry one can proceed along similar lines. The null geodesics of interest (emanating from $(\varphi_0,0)$) are given by:
\begin{eqnarray}
\label{tphgeodBTZ}
t(r) &=&  \varphi_0 +  \frac{1}{2 r_+} \, \ln
\frac{\sqrt{(1-j^2) \, r^2 + j^2 \, r_+^2}
- r_+ }{\sqrt{(1-j^2) \, r^2 +j^2 \, r_+^2} + r_+ } \\
\varphi(r) &=& \frac{1}{2 r_+} \, \ln \frac{\sqrt{(1-j^2) \, r^2 + j^2 \, r_+^2} + j \, r_+}{
\sqrt{(1-j^2) \, r^2 + j^2 \, r_+^2} - j \, r_+}
\end{eqnarray}
which determines $\bcwedgef$. Again using the symmetries we realize the the past and future Rindler horizons must intersect at $t=0$. Then setting $t=0$ above and solving for $j =  \coth (r_+ \varphi_0) \, \tanh (r_+ \varphi)$  then leads to the desired co-dimension two surface:
\begin{align}  \label{csfbtz}
&\text{(b).}&  \csf{\cal A}: && t & = 0\,, 
&r^2 (\varphi) = r_+^2 \
{ \cosh^2 ( r_+ \, \varphi_0)  \over \sinh^2
  ( r_+ \, \varphi_0) \, \cosh^2  ( r_+ \, \varphi) - \cosh^2
   ( r_+ \, \varphi_0) \, \sinh^2  ( r_+ \, \varphi) }
\end{align}
which as advertised agrees with the extremal surface \eqref{extrbtz}.

\noindent
{\bf (c).} Finally, for the rotating BTZ spacetime \eqref{rbtz} we use the mapping to the Poincar\'e coordinates \eqref{wzdefs}. Since the boundary domain of dependence is still given by \eqref{domads3} we simply need to determine the light cones from the points $(w_\pm,z) =(e^{\pm\, (r_+\pm ,r_-) \varphi_0} ,0) $ which is easily done since the geometry \eqref{pads3} is conformally flat. The boundary of the causal wedge in the bulk is given by
\begin{equation}  \label{rphrbtz}
\bcwedgef: \qquad \left(w_+ - e^{(r_++r_-)\varphi_0} \right) \left(w_- - e^{(r_+-r_-)\varphi_0} \right) + z^2 =0
\end{equation}
and $\bcwedgep$ is simply obtained by replacing $\varphi_0 \to -\varphi_0$ above. The surface $\csf{\cal A}$ lies at the intersection of $\bcwedgef$ and $\bcwedgep$ which in particular means that it lies on a  co-dimension one hypersurface in the bulk
\begin{align} 
&\text{(c)}.& \sinh(r_+-r_-)\, w_+ - \sinh(r_+-r_-) \, w_- = -\sinh(2\,r_-) \,,\qquad 
\label{csfrbtz}
\end{align}
which is indeed the boosted surface \eqref{extrrbtz} on which the extremal surface $\extr{\cal A}$ lies. One can further go on to show that $\csf{\cal A}$ in fact coincides with $\extr{\cal A}$. As is clear from the plot in \fig{f:cwedge3d}(c), in the non-static case the surfaces $\csf{\cal A} = \extr{\cal A}$ do not lie on a fixed time slice, even when ${\cal A}$ lies on one at the boundary.

In all three geometries we thus see that the extremal surface relevant for entanglement entropy coincides with the causally motivated surface. To a certain extent this is to be expected for the black hole spacetimes given that the surfaces coincide for the pure \AdS{3} spacetime, owing to the fact that the latter are locally \AdS{3}. Given this information, we can also conclude from previous computations that:
\begin{align}
\text{(a).}& \qquad  S_{\cal A} = \chi_{\cal A} = \frac{c_\text{eff}}{3}\, \log\left(\frac{2\varphi_0}{\varepsilon}\right) \\
\text{(b).}& \qquad   S_{\cal A} = \chi_{\cal A} = \frac{c_\text{eff}}{3}\, \log\left[ \frac{\beta}{\pi\, \varepsilon}\,
\sinh\left(\frac{2\pi\,\varphi_0}{\beta}\right) \right] \\
\text{(c).}& \qquad  S_{\cal A}= \chi_{\cal A} = \frac{c_\text{eff}}{6}\, \log\left[ \frac{\beta_+\,\beta_-}{\pi^2\, \varepsilon^2}\,
\sinh\left(\frac{2\pi\,\varphi_0}{\beta_+}\right) \sinh\left(\frac{2\pi\,\varphi_0}{\beta_-}\right) \right] 
\end{align}

Before proceeding further with the discussion we should note that the agreement between $S_{\cal A}$ and $\chi_{\cal A}$ does not extend to other states or density matrices of $1+1$ dimensional CFTs. The general argument from a holographic perspective was motivated above and has previously been given more explicitly in \cite{Hubeny:2007xt}, which we simply quote here without further proof. Given that static rotationally symmetric states of a $1+1$ CFT on a cylinder are dual to static asymptotically \AdS{3} geometries of the form:
\begin{equation}
ds^2 = -f(r)\, dt^2 + h(r)\, dr^2 + r^2\, d\varphi^2
\label{metgsss}
\end{equation}	
If we are interested in the entanglement entropy, then we simply compute the area of a minimal surface at a constant $t$ slice, in particular noting that such a minimal surface is insensitive to the redshift function $f(r)$. On the other hand, the causal construction requires us to construct light-cones in the bulk spacetime which care about the metric functions (up to an overall conformal factor, which we can gauge-fix to be $r^2$). It then follows that the surface $\csf{\cal A}$ for a given region ${\cal A}$ cares about the redshift factor. 
More specifically, the minimal radius reached by  $\extr{\cal A}$ is given by the conserved angular momentum $J$ along the $\varphi$ direction, while the minimum radius reached by the causal surface $\csf{\cal A}$ depends on both $f(r)$ and $h(r)$. In particular, in order for $\extr{\cal A}$ and $\csf{\cal A}$ to coincide, the spacetime \req{metgsss} would minimally need to satisfy 
\begin{equation}
\int_\ro^{\infty} \frac{\ro \, \sqrt{h(r)}}{r \, \sqrt{r^2 - \ro^2}} \, dr 
	= \int_\ro^{\infty} \frac{\sqrt{h(r)}}{\sqrt{f(r)}} \, dr
\label{eqcond}
\end{equation}	
where the LHS is an expression for the angle $\varphi_0$ reached by a constant-$t$ spacelike geodesic which passes through $\ro$ at $t=0, \varphi=0$, whereas the RHS corresponds to the time at which a radial null geodesic at $\varphi=0$ which starts from $r=\ro$ at $t=0$ reaches the boundary $r=\infty$ -- this would be $\ddtipf$ for $\domdy^+[{\cal A}]$ with ${\cal A}= \varphi \in \{ - \varphi_0, \varphi_0 \}$, so that $t(\ddtipf) = \varphi_0$.
Note that \req{eqcond} is automatically satisfied for $f(r) = \frac{1}{h(r)} = r^2 + \alpha$ for any $\alpha$ -- as demonstrated above for AdS$_3$ and BTZ; however, it is certainly not true in full generality.  For example one can easily check that \req{eqcond}  is not satisfied for e.g.\ $f(r) = \frac{1}{h(r)} = r^2 + 1 - \frac{\alpha}{r^2}$, just to pick a random example. This was already explained in Appendix A.2 of \cite{Hubeny:2007xt}.

\subsubsection{Spherical entangling surfaces and vacuum state of CFT$_d$}
\label{s:}

Our next set of examples concerns conformal field theories in $d > 2$. The theories are taken to live either in Minkowski spacetime ${\mathbb R}^{d-1,1}$ or on the Einstein Static Universe  (ESU) ${\bf S}^{d-1} \times {\mathbb R}$. The regions of interest are a spherical ball in flat space and a slice of the sphere at constant latitude in the ESU:
\begin{align}
(i). \qquad { \cal B}&= {\mathbb R}^{d-1,1}: & {\cal A} & = \left\{ (t,\vec{x} ) \mid t = 0 ,\; \rho^2 \leq a^2 \right\}, \qquad \vec{x} = \{ \rho , \Omega_{d-2}\} 
\label{ddisc}\\
(ii). \qquad {\cal B}&= {\bf S}^{d-1} \times {\mathbb R}: & {\cal A} & = \left\{ (t, \theta, \Omega_{d-2} ) \mid t = 0 ,\; \mid\!\theta\! \mid\leq \theta_0 \right\} 
\label{dsph}
\end{align}
where we find it convenient to introduce polar coordinates on ${\mathbb R}^{d-1}$. The dual geometries for the two cases are of course Poincar\'e \AdS{d+1} and the global \AdS{d+1} respectively, for which we use the metrics:
\begin{align}
(i). &&  ds^2 & = \frac{-dt^2 +dz^2 +  d\rho^2 + \rho^2\, d\Omega_{d-2}^2 } {z^2} 
\label{pflat}\\
(ii). &&  ds^2 & = -\left(1+r^2 \right)dt^2 +\frac{dr^2}{1+r^2} +  r^2\, \left(d\theta^2 + \sin^2\theta \, d\Omega_{d-2}^2 \right) 
\label{gads}
\end{align}
For these cases the causal structure of the boundary is simple and one can easily ascertain the domain of dependence $\domd$. In the flat spacetime example, we simply need the cone over the spherical ball,  while in the ESU we can use the flat metric in the two dimensional $(t,\theta)$ plane to chart out the causal development.
\begin{align}
(i). && \domd & = \left\{ (t,\rho, \Omega_{d-2} ) \mid t \ge 0\,,t+ \rho \leq a \; \cup \;  t \le 0\,,  \rho -t \leq a \,,\;\;   \Omega_{d-2}\; \text{arbitrary} \right\}
\label{domdisc}
 \\
(ii). &&  \domd & = \left\{ (t,\theta, \Omega_{d-2} ) \mid t  \leq  |\theta_0 - \theta| \,, \theta \in ( 0,\theta_0) 
\;\cup \;  t \leq |\theta_0 +\theta| \,, \theta \in (-\theta_0, 0) , \Omega_{d-2}\; \text{arbitrary} \right\}
\label{domsph}
\end{align}

Once again let us recall the computation of entanglement entropy in these cases. The minimal surface prescription can be easily used here to compute $S_{\cal A}$ \cite{Ryu:2006ef}; for case (i) it is simply a hemisphere, while for the ESU we will be able to explicitly construct the geodesic congruences of interest by integrating the geodesic equations using the $SO(d-1)$ symmetry at our disposal. The results are 
\begin{align}
&(i).& \extr{\cal A}:  && t & = 0\,, & z^2 + \rho^2 &= a^2 \,, \qquad  \Omega_{d-2} \; \text{arbitrary}
\label{extrflat} 
\\
&(ii).& \extr{\cal A}: && t&=0\,, & r^2 (\theta) &= \frac{\cos^2 \theta_0 }{ \sin^2 \theta_0 \, \cos^2 \theta - \cos^2 \theta_0 \, \sin^2 \theta } \,, \qquad  \Omega_{d-2} \; \text{arbitrary}
 \label{extresu}
\end{align}	

More importantly, as originally deduced in \cite{Myers:2010tj} and subsequently elaborated in \cite{Casini:2011kv}, these two situations are special. In these circumstances the reduced density matrix $\rho_{\cal A}$ can be mapped by a unitary transformation into a thermal density matrix. The argument in fact relies on the fact that the interior of the domain of dependence $\domd$ in the two cases can be mapped to a hyperbolic cylinder ${\mathbb H}_{d-1} \times {\mathbb R}$ by a conformal transformation (or equivalently to the static patch of de Sitter spacetime dS$_{d}$). The same conformal transformation acts as a unitary transform on the density matrix; its  effect is simply to convert the reduced density matrix into a thermal one. 

Once we have converted the computation of the entanglement entropy to the computation of a thermal partition function we have simplified things considerably. Not only can one do the computation of the latter in free theories (see for example 
\cite{Klebanov:2011uf}), but in the holographic context one can use the knowledge of the thermal state for a CFT on ${\mathbb H}_{d-1} \times {\mathbb R}$ being given by the hyperbolic black hole as described originally in \cite{Emparan:1999gf}. This strategy was used in \cite{Casini:2011kv} to prove the minimal surface holographic entanglement entropy proposal of \cite{Ryu:2006bv}. The essential point of note is that the conformal map on the boundary, which corresponds to a bulk diffeomorphism, takes the minimal surface relevant for the entanglement entropy computation onto the bifurcate Killing horizon of the hyperbolic black hole. The bifurcation surface of the Killing horizon being directly relevant for the computation of the thermal entropy of the field theory on the hyperbolic cylinder, one can clearly see that the the entanglement entropy in the original conformal frame is captured by the  extremal surface.

Having understood the entanglement entropy in these cases, let us tackle the causal holographic information. We have already  constructed the boundary domain of dependence for the regions of interest in \eqref{domdisc} and \eqref{domsph}. We now need the bulk causal wedge associated to these regions in the geometries \eqref{pflat} and \eqref{gads} respectively.

\noindent
{\bf (i).} Let us start with the CFT on ${\mathbb R}^{d-1,1}$; using the conformal flatness of  the metric \eqref{pflat} we ignore the overall $z^{-2}$ and just  write down the bulk light-cone emanating from the top tip of the boundary cone $\domd$, i.e., from $ t = a\,,\rho=0$. This immediately leads to:
\begin{align}
\label{rphpoin}
\bcwedgef: \qquad 
\left\{ (t,\rho,\Omega_{d-2},z) \mid \, z =\sqrt{\left(a-t\right)^2 + \rho^2} , \;\Omega_{d-2} = \text{arbitrary}\right\}
\end{align}
Similarly the future Rindler horizon from the bottom tip of the cone i.e., from $ t = -a\,,\rho=0$ is obtained by $t \to -t$ above. The causal wedge $\cwedge$ itself is the entire bulk region sandwiched between the past and future Rindler horizons and the boundary. With this information  at hand, it is easy to find the bulk co-dimension two surface relevant for the causal holographic information. This is simply the surface at $t =0$ which lies and the intersection of the past and future Rindler horizons and is easily seen to be a  hemisphere:
\begin{align}
&(i).& \csf{\cal A}:  && t & = 0\,, & z^2 + \rho^2 &= a^2 \,, \qquad  \Omega_{d-2} \; \text{arbitrary}
\end{align}

\noindent
{\bf (ii).} We now turn to the case of the CFT on the ESU and a region that is a segment of the spatial sphere. The domain of dependence on the boundary is given by \eqref{domsph} and the causal construction of the bulk region proceeds along the same lines as outlined in the global \AdS{3} example earlier. In particular, we simply look for past/future null geodesic congruences emanating from $t = \pm\theta_0$ which are $SO(d-1)$ symmetric. These congruences can be easily understood in the three dimensional geometry spanned by constant $\Omega_{d-2}$ sections of \eqref{gads}. The geodesic congruence from the top $t = \theta_0$ spans the surface 
\begin{equation} 
\bcwedgef:  r^2(t,j) = \frac{\cot^2 \left(\theta_0-t\right)+ j^2}{1 -j^2} \ , \quad
\theta(t,j) = \frac{\pi}{2} - \tan^{-1} \left(\frac{\cot \left(\theta_0-t\right)}{j} \right) \ , \quad 
\Omega_{d-2}  = \text{arbitrary}
\end{equation}
where $j$ parameterizes the different curve in the congruence. From this expression, 
we can see that the causal holographic surface of interest is simply:
\begin{align} 
&(ii).&  \csf{\cal A}: && t & = 0\,,
& r^2 (\theta) =\frac{\cos^2 \theta_0}{\sin^2 \theta_0 \, \cos^2 \theta - \cos^2 \theta_0 \, \sin^2 \theta}
\,, \qquad  \Omega_{d-2} \; \text{arbitrary}
 \label{csfgads}
\end{align}
which indeed agrees with the  extremal surface \eqref{extresu}.
 
\subsubsection{Relation between entanglement and causal holographic information}
\label{s:rel}

We have thus far seen that while $\chi_{\cal A}$ does not generically qualify to be an entropy, it curiously agrees with the holographic entanglement entropy in all cases where there is a microscopic understanding/derivation of the latter. We conjecture that this concordance is a result of these specific states and regions capturing the maximal information available in the field theory and therefore $S_{\cal A}$ contains as much information as $\chi_{\cal A}$.

To push this further, let us also highlight another example wherein the two constructions agree.\footnote{
We thank Roberto Emparan for pointing this out to us.} Consider a field theory living on a black hole background. We can either imagine this in the brane-world context, where we have induced gravity coupled to the field theory as discussed in \cite{Emparan:2006ni}, or simply let the black hole be a rigid gravitational background for  the field theory as described for instance in \cite{Hubeny:2009ru}. To keep things simple we shall for the moment demand that the field theory in the region outside the black hole is in its vacuum state; this would be the analog of the Boulware vacuum discussed in the case of asymptotically flat black holes. The bulk duals for such Boulware states of four dimensional holographic field theories in the Schwarzschild background have been numerically constructed in \cite{Figueras:2011va}. 

 Since the background geometry on which the field theory lives has a black hole metric, there is natural region which we can consider, viz., the region interior to the brane-world/boundary black hole. The field theory degrees of freedom on either side of the horizon are entangled with each other and this entanglement entropy can be easily shown to be captured by the area of the bulk black hole horizon that ends on the brane/boundary \cite{Emparan:2006ni} (such bulk black holes were called black droplets in \cite{Hubeny:2009ru}). The key point to note is that the bulk horizon is by itself an extremal surface; thus establishing a link between entanglement and black hole entropies in these induced gravity set-ups. Furthermore, since we have taken the field theory outside to be in the vacuum state, it also follows that the entanglement entropy of the spatial region outside the black hole is given by the area of the bulk black hole horizon in Planck units. This will be the region ${\cal A}$ we focus on.
 
 For the purposes of our discussion all we need to know is the fact that the spatial region outside the black hole horizon on the boundary has as its causal wedge the exterior region of the bulk black hole spacetime. In particular, in this example the boundary of the bulk causal wedge  $\cwedge$ is simply the black hole horizon in the bulk geometry. From this it follows that the causal extremal surface of interest is the bulk bifurcation surface and indeed $\chi_{\cal A}$ is also given by the bulk black hole horizon area. 
Hence we once again have a situation in which the causal holographic information agrees with the entanglement entropy. Note that this was predicated upon our insisting that the state of the field theory in the region outside the boundary black hole is the vacuum state. One could have considered a density matrix in the region outside, say for example the thermal density matrix, but that would give rise to more complicated entanglement. The bulk dual for such situations is also more involved, for it contains either a droplet and a second bulk horizon or a single connected horizon in the bulk (the black funnel), depending on the characteristics of the black hole \cite{Hubeny:2009ru}.

The moral of our handful of examples seems to be that when the degrees of freedom in ${\cal A}$ are maximally entangled with the degrees of freedom in the complement ${\cal A}^c$ one ensures that $\chi_{\cal A} = S_{\cal A}$. To flesh this out further, let us imagine for a moment that the boundary theory can be viewed as a quantum system with a finite dimensional Hilbert space associated with the region ${\cal A}$. Since the region is entangled with its complement ${\cal A}^c$ we have to ask how best can we entangle the degrees of freedom in ${\cal A}$ with those in ${\cal A}^c$ and thus maximize the entropy of the 
reduced density matrix $\rho_{\cal A}$. For the finite dimensional gedanken system under consideration it is clear that we can at most have an entropy given by the thermal entropy. The best entanglement is achieved when we correlate degrees of freedom in ${\cal A}$ as we would in the construction of the thermofield double of the Hilbert space associated with ${\cal A}$. Specifically, it does not matter if we can enlarge the system ${\cal A}^c$ arbitrarily; entanglement is saturated when ${\cal A}^c$ is similar in size to ${\cal A}$. 

Let us now turn from such special situations to the generic case.   For arbitrary (physically sensible) state and arbitrary region ${\cal A}$, we conjecture that we can bound\footnote{
Here we use the terminology {\it bound} in the sense of an inequality between two quantities, both of which depend on the state, rather than in the more typical sense of one variable quantity always being smaller than a pre-specified number.
} $S_{{\cal A}}$ by $\chi_{\cal A}$, and moreover that the causal holographic information surface  $\csf{\cal A}$ lies outside of (i.e.\ closer to the boundary than) the entanglement entropy (extremal) surface $\extr{\cal A}$.  
We can understand these statements as follows.

In a static geometry, both $\csf{\cal A}$ and $\extr{\cal A}$ lie on the same time slice by symmetry, and moreover, out of all surfaces on that time slice, $\extr{\cal A}$ is defined to be the one with {\it minimal} area, so by definition  $S_{{\cal A}} \le \chi_{\cal A}$.  For general time-dependent configurations, the argument is not as straightforward, but we nevertheless expect that in physically reasonable situations this inequality will continue to hold.  

\begin{figure}
\begin{center}
\includegraphics[width=6.in]{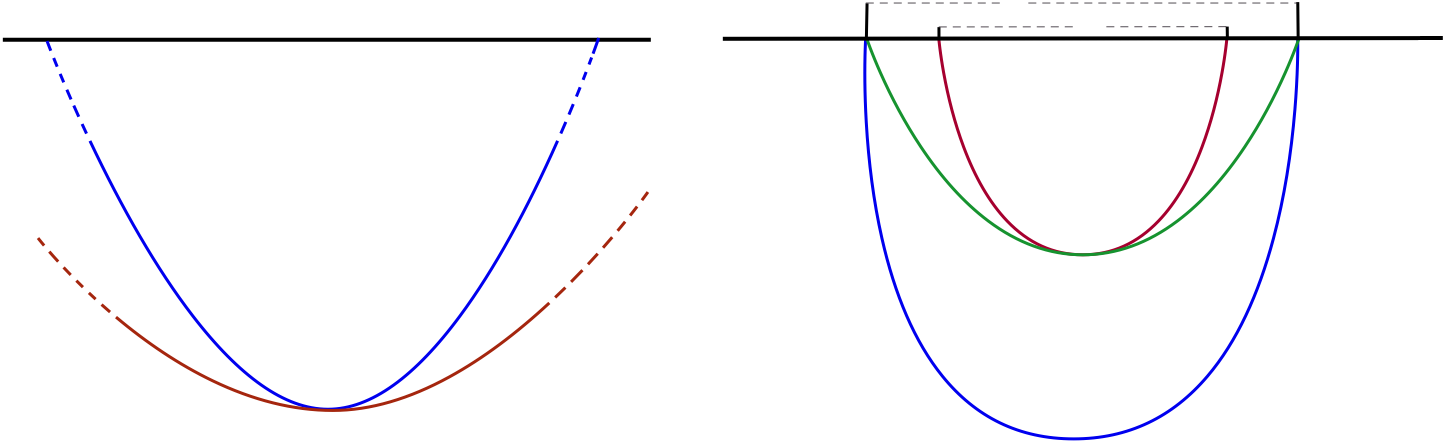}
\begin{picture}(1,1)(0,0)
\put(-275,128){\makebox(0,0){$\text{bdy}$}}
\put(-337,9.5){\makebox(0,0){\small{$\bullet$}}}
\put(-335,2){\makebox(0,0){$p$}}
\put(-262,85){\makebox(0,0){${\cal S}_\alpha$}}
\put(-260,40){\makebox(0,0){${\cal S}_\beta$}}
\put(-20,128){\makebox(0,0){$\text{bdy}$}}
\put(-75,5){\makebox(0,0){$\csf{\cal A}$}}
\put(-80,57){\makebox(0,0){$\extr{\cal A}$}}
\put(-138,90){\makebox(0,0){$\csf{{\tilde{\cal A}}}$}}
\put(-110,128){\makebox(0,0){${\tilde{\cal A}}$}}
\put(-135,133){\makebox(0,0){${\cal A}$}}
\put(-112,55.5){\makebox(0,0){\small{$\bullet$}}}
\put(-112,48){\makebox(0,0){$p$}}
\end{picture}
\caption{
Sketch accompanying the argument in main text for why the extremal surface  $\extr{\cal A}$ cannot lie closer to the boundary than the causal information surface $\csf{\cal A}$.
Left: we argue that at $p$, $\Theta_{\alpha} < \Theta_{\beta}$.  Right: impossible situation, since it contradicts the physical requirement that $\Theta_{\csf{\tilde{{\cal A}}}} \ge 0$ and $\Theta_{\extr{\cal A}} = 0$ everywhere.}
\label{f:expansions}
\end{center}
\end{figure}
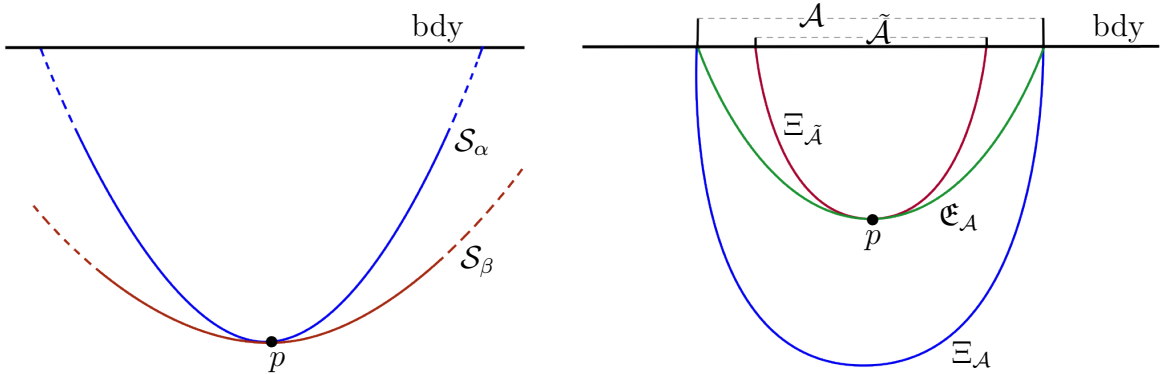

We now sketch an argument in support of this assertion.\footnote{
Related observations have been made independently by Aron Wall.}  
We first establish a relation between the relative position of two surfaces anchored on the same boundary region $\partial{{\cal A}}$ and the expansion $\Theta$ along null normals to these surfaces.  
Specifically, $\Theta$ is the expansion of the null geodesic congruence emanating from the surface of interest and we are only interested in the outgoing congruence (either future/past directed), i.e., the congruence that reaches the boundary (or terminates in a caustic along the way).

We start by observing that given two  surfaces, ${\cal S}_\alpha$ and ${\cal S}_\beta$ which are tangent at some point $p$ as shown in left panel of \fig{f:expansions}, such that ${\cal S}_\alpha$ is more bent towards the `outward' direction, the expansion of ${\cal S}_\alpha$ at $p$ must be smaller than that of ${\cal S}_\beta$, 
\begin{equation}
\Theta_{\alpha} < \Theta_{\beta} \ ,
\label{expreln}
\end{equation}	
since the bending makes the null normals converge more.  
Using this observation, we proceed to construct a proof by contradiction.  Suppose we had a situation where $\csf{\cal A}$ was located further from the boundary than $\extr{\cal A}$, as in the right panel of \fig{f:expansions}.  Then there must exist a boundary region  $\tilde{{\cal A}} \subset {\cal A}$ such that the causal information surface corresponding to this smaller region $\csf{\tilde{{\cal A}}}$ just touches $\extr{\cal A}$, i.e. is tangent at some point $p$.
Since $\csf{\tilde{{\cal A}}}$ by definition lies on the boundary of the causal wedge 
$\blacklozenge_{\tilde{{\cal A}}}$
 (so that the null normals must reach the boundary -- i.e.\ must extend to infinite affine parameter without encountering caustics), the causal information surface $\csf{\tilde{{\cal A}}}$ must have non-negative expansion, $\Theta_{\csf{\tilde{{\cal A}}}} \ge 0$.  But that would force the expansion for an extremal surface to be strictly positive at the point where the two surfaces are tangent, $\Theta_{\extr{\cal A}}(p) > 0$ by \req{expreln}.  This is a contradiction, since by construction,  the expansion for an extremal surface must vanish everywhere, as proved in  \cite{Hubeny:2007xt}.  Hence we conclude that in order for $\extr{\cal A}$ to be an extremal surface (i.e.\ $\Theta_{\extr{\cal A}} = 0$) and $\csf{\cal A}$ to be a causal holographic information surface (i.e.\ $\Theta_{\csf{\cal A}} \ge 0$), the causal holographic information surface $\csf{\cal A}$ must either lie closer to the boundary than (or at best coincide with) the extremal surface $\extr{\cal A}$.
 
This solidifies our expectation that the causal information cannot be any smaller than the entanglement entropy, since the former corresponds to the area of a surface located closer to the boundary where the warp factor is larger, leading to greater area, $S_{{\cal A}} \le \chi_{\cal A}$.

\section{Discussion}
\label{s:discuss}

The primary issue that motivated the present discussion is how much information is there about the bulk in a given spatial region of the field theory. To make this question sharper, we  imagine that one has access to a suitable algebra of observables of the field theory localized to the region ${\cal A}$ in question and that one is also equipped with the knowledge of the reduced density matrix $\rho_{\cal A}$. Armed with this information,  one can predict the quantum development of $\rho_{\cal A}$ in the boundary domain of dependence $\domd$. The question then is: what part of the bulk should one hope to reconstruct given such data on the boundary? 

As we have argued, a very natural region from the bulk standpoint is the   bulk causal wedge $\cwedge$ associated with the region ${\cal A}$. This region is composed of all points in the bulk which can both influence and be influenced by some part of  $\domd$, the latter being the boundary region which is fully determined by ${\cal A}$.  The causal wedge $\cwedge$ is therefore defined purely by causal relations, and we believe that this inherent simplicity translates to correspondingly natural (albeit perhaps not as readily apparent) construct in the dual field theory. It seems natural to expect that $\cwedge$ gives the minimal spacetime volume that should be reconstructable from the data contained in ${\cal A}$, since we can imagine `observers' sent from the field theory being able venture into this region and returning to the boundary within $\domd$. More specifically, we are suggesting that the reduced density matrix $\rho_{\cal A}$ (which is more naturally associated with the full $\domd$ rather than just ${\cal A}$)  can be used to recover the bulk geometry  at least in $\cwedge$. As we have stressed at various points, it may be possible to reconstruct more of the bulk geometry; here we have proposed what we think is the conservative option. 

In the bulk, the causal wedge  is bounded by two null surfaces, whose intersection is a bulk co-dimension two surface which we dubbed  causal information surface, denoted by $\csf{\cal A}$.  This is a special surface within $\cwedge$: it reaches deepest into the bulk and has extremal area among all surfaces on $\bcwedge$.  It is also anchored on our entangling surface $\partial {\cal A}$ on the boundary.   We proposed to associate the area $\chi_{\cal A}$ of the causal information surface $\csf{\cal A}$ (measured naturally in Planck units analogously to the Bekenstein-Hawking formula)  to a certain measure of the information contained in ${\cal A}$ -- hence the name {\it causal holographic information}.	 We  argued that $\chi_{\cal A}$ serves to bound the entanglement entropy $S_{\cal A}$ of the reduced density matrix from above. Nevertheless, $\chi_{\cal A}$ is not a von Neumann entropy, as it  fails to satisfy the sub-additivity condition. However, in certain special circumstances where the degrees of freedom in ${\cal A}$ are maximally entangled with those outside, we have further shown that $\chi_{\cal A}$ agrees with the entanglement entropy.
 
These features then appear to suggest that $\chi_{\cal A}$ should be viewed as a lower bound on the information about the bulk spacetime carried by the given region. In  other words, it is natural to conjecture that the minimal information contained in the region ${\cal A}$ is measured by $\chi_{\cal A}$. At first this seems somewhat counter-intuitive, since we have said that $\chi_{\cal A} \ge S_{\cal A}$. So why is it a lower bound on the `information content'?\footnote{
In interest of clarity let us remark that we are being necessarily vague about the notion of information here. To flesh this out more would require a detailed specification of the reconstruction procedure itself, which is of course a notoriously difficult problem.} 
To understand this point, let us note that there exist situations where field theory observables are sensitive to a larger region of the bulk spacetime (assuming a suitable notion of analyticity in the bulk).\footnote{ The importance of assuming analyticity was stressed in particular in \cite{Louko:2000tp}. 
In analytic spacetimes it is possible to approximate the correlation functions via a suitable geodesic approximation.} For instance, field theory correlation functions, even with the operator insertions restricted to lie in $\domd$, can be sensitive to the bulk outside the causal wedge when we evaluate them using a geodesic approximation. Similarly, the entanglement entropy is computed by the area of an extremal surface which typically  reaches deeper into the bulk (which motivates the inequality $S_{\cal A} \leq  \chi_{\cal A}$). We then seem to be missing information about such regions outside the causal wedge when we characterize the information by $\chi_{\cal A}$ and thus it seems natural to think of $\chi_{\cal A}$ as providing a lower bound on the information contained within a given region on the boundary.

There is an additional subtlety which enters our considerations. While it seems natural to ask which surface reaches deeper into the bulk, we should be aware that statements such as $\chi_{\cal A} \geq S_{\cal A}$ should be interpreted with care owing to the fact that both quantities are divergent. Also from explicit computations of \sec{s:connections} it is clear that in some simple cases $\chi_{\cal A}$ is divergently larger than $S_{\cal A}$ (see \eqref{entchistrip}). One might be tempted to make the comparison more meaningful by suitably regulating the two quantities.  However, this is not as straightforward as one might imagine.\footnote{
 We thank Hong Liu for useful discussions on this issue; see also \cite{Liu:2012ee}.}  The most obvious way to regulate the areas is by background subtraction. Given ${\cal A}$ and some state (or more generally a density matrix) of the field theory $\ket{\psi}$ we can consider $S^{\text{reg}}_{\cal A}= S_{\cal A}(\ket{\psi}) - S_{\cal A}(\ket{0})$ and $\chi^{\text{reg}}_{\cal A}= \chi_{\cal A}(\ket{\psi}) - \chi_{\cal A}(\ket{0})$, where we have taken the vacuum state $\ket{0}$ as our reference state to perform the subtraction. The utility of such a subtraction is that we should be rid of the divergent parts of the two quantities. A-priori it is not clear that $\chi^{\text{reg}}_{\cal A}$  will continue to be larger than $ S^\text{reg}_{\cal A}$. In fact, preliminary investigations of certain simple cases seem to suggest that $\chi^\text{reg}_{\cal A} < S^\text{reg}_{\cal A}$. If this were to be established more firmly one could view it as more intuitive support of the speculation that $\chi_{\cal A}$ is the minimal information available in the field theory.

One should perhaps emphasize that in the above discussion, $\chi_{{\cal A}}$ served merely to quantify the amount of information in ${\cal A}$ -- it did not itself specify the actual information.
However, we can elevate our construction to extract more useful information about the bulk geometry.  In particular, instead of considering just a single quantity, $\chi_{{\cal A}}$, we can imagine that we have access to infinitely many, more refined, quantities, namely $\chi_{{\cal Q}}$ corresponding to all subregions ${\cal Q} \subset {\cal A}$.  It is then an interesting question whether the knowledge of this particular set of quantities allows us to reconstruct the full bulk geometry within $\cwedge$. 
For static spherically symmetric spacetimes, the answer is almost certainly yes, since the amount of information to recover (namely two functions of one variable) is much smaller than the information available in $\{ \chi_{{\cal Q}} \}$ (one function of several variables).  This naive counting would in fact suggest that 
$\{ \chi_{{\cal Q}} \}$ should provide access to the bulk geometry within $\cwedge$ for more general configurations as well; we however leave the explicit reconstruction as an interesting problem for the future.

Let us also note that there is a sense in which the knowledge of $\{ \chi_{\cal Q} \}$ with ${\cal Q} \subset {\cal A}$ provides an easier way to reconstruct the geometric data of the bulk than by using, say, the entanglement entropy. For simplicity, let us consider static bulk spacetimes. If we use the entanglement entropy of the reduced density matrix $\rho_{\cal A}$, and restrict to the conventional Hamiltonian evolution in $\domd$, then it is not easy to reconstruct the time component of the metric (the redshift factor). This is because the bulk dual of $S_{\cal A}$ is  characterized by extremal surfaces which are localized on a single time slice in the bulk and therefore are insensitive to $g_{tt}$. On the other hand, since the causal construction depends on the full bulk metric, it in particular depends on the time component as well. Naively, in constructing $\cwedge$ we seem to miss one degree of freedom, the conformal factor, since for purposes of discussing causal relation we can work in a conformally rescaled spacetime. However, in our definition of $\chi_{\cal A}$ we reinstate the conformal factor, since we are required to measure the proper spacetime area of $\csf{\cal A}$. Thus armed with knowledge of $\chi_{\cal A}$ we could in principle extract the bulk metric more easily within $\cwedge$. 

Finally, it is worth emphasizing that there are two distinct notions pertaining to `bulk information' that we have been discussing, and while it is tempting to conflate the two, it is a-priori not clear that they are equivalent. On the one hand, we have talked about actual reconstruction of the bulk spacetime, and on the other, the ability to probe the bulk.  By the latter we simply mean the existence of a CFT observable which is sensitive to the bulk physics, though it may not carry enough information to fully determine the physics. We believe that it should be possible to reconstruct more than the causal wedge, though how much more is unclear. It would be interesting to ask how to use the ideas presented herein to actually carry out the reconstruction and to understand the utility of the causal construction in greater detail; we hope to report on these issues in the future.

\acknowledgments 
It is a pleasure to thank Roberto Emparan, Gary Horowitz, Juan Maldacena,  Don Marolf, Hong Liu, Mark van Raamsdonk, Mark Srednicki, and Tadashi Takayanagi  for extremely useful discussions. We would also like to thank Hong Liu, Don Marolf, Tadashi Takayanagi and Mark van Raamsdonk for their comments on an early draft of the paper.
 MR would like to thank the organizers of the Iberian strings meeting in Bilbao and the Indian-Israeli strings meeting in Jerusalem for hospitality during the course of this project. MR and VH would also like to thank the KITP for hospitality during the ``Bits, branes and black holes'' workshop. This research was supported in part by the National Science Foundation under Grant No. NSF PHY11-25915 and by the STFC Consolidated Grant ST/J000426/1.
\appendix
\section{Caustics for higher-dimensional regions ${\cal A}$}
\label{s:caustics}

In 1+1 dimensional CFT, the region ${\cal A}$ is an interval and therefore the domain of dependence $\domd$ has a single future tip $\ddtipf$ and single past tip $\ddtipp$, as apparent from the left panel of \fig{f:csets}.
This fact makes the definition of the corresponding bulk causal wedge $\cwedge$ particularly convenient, as its boundary  $\bcwedge$ is expressible in terms of two Rindler horizons.  This convenient characterization however does not generically carry over to higher dimensions.  For an arbitrary region ${\cal A}$ in 2 or more spatial dimensions, the future and past boundary domains of dependence $\domdy^\pm[{\cal A}]$ will not end by a single tip, but rather a co-dimension one set, which we refer to as {\it caustic}.  
Here we explore these caustics a bit more, in the case of $d=3$.

Let us consider $2+1$ dimensional boundary, so that the region ${\cal A}$ is 2-dimensional, and correspondingly its boundary (also referred to as the entangling surface) $\entsurf$ is a closed curve in $\RR^2$.  Let us parameterize it by $\lambda \in [0,1]$, and denote
\begin{equation}
\entsurf(\lambda) = (x(\lambda), y(\lambda))
\label{}
\end{equation}	
where $x(0)= x(1)$ and  $y(0)= y(1)$.  
Now we want to characterize the boundary of the future and past domains of dependence of ${\cal A}$, in order to specify $\domd$.  Since the boundary spacetime is static, $\domdy^+[{\cal A}]$ and $\domdy^-[{\cal A}]$ are symmetric to each other, so it suffices to consider just $\domdy^+[{\cal A}]$.
Being as causal set, its boundary is generated by future-directed null geodesics from $\entsurf$ which emanate orthogonally to $\entsurf$.  The locus of points where these null geodesics intersect is called a caustic.  Such a caustic terminates the boundary of $\domd$, since further continuation of these null geodesics does not lie within $\domd$.  Moreover, we can in fact characterize $\domd$ by specifying the set of all caustics $\caustic^+$ 
 (which by symmetry also determines $\caustic^- $), 
as 
\begin{equation}
\domd = I_{\partial}^-[\caustic^+] \cap  I_{\partial}^+[\caustic^-] \ .
\label{}
\end{equation}	
 In general, a caustic for a 2-dimensional null surface will occur along a 1-dimensional spacelike curve, which may branch and terminate.  For the special case of ${\cal A}$ being a round disk, these caustics degenerate to a single point $\ddtipf$.  However, any other shape of ${\cal A}$ will have more than a single point in $\caustic^\pm$.

To determine $\caustic^\pm$ explicitly for a given region ${\cal A}$, we proceed as follows.
For $\lambda$ increasing in anti-clockwise manner, the inward-pointing spatial normal to $\entsurf$ of unit norm at each $\lambda$ is given by 
\begin{equation}
\psi(\lambda) = \frac{1}{n(\lambda)} \, \left( - y'(\lambda), x'(\lambda) \right)
	\ , \qquad {\rm where} \ \ 
	n(\lambda) = \sqrt{x'(\lambda)^2 + y'(\lambda)^2}
\label{}
\end{equation}	
which yields the following generators, parameterized by $t$ and labeled by $\lambda$:
\begin{equation}
k_{\lambda}(t) =  \left( t \, , \, 
x(\lambda) - t \, \frac{y'(\lambda)}{n(\lambda)}  \, , \, 
y(\lambda) + t \, \frac{x'(\lambda)}{n(\lambda)}  \right)
\label{}
\end{equation}	
The caustics occur when $k_{\lambda_1}(t) = k_{\lambda_2}(t) $ for some $\lambda_1$, $\lambda_2$ and $t$.  While this is difficult to calculate in general, we can easily find caustics in specific examples.  
\begin{figure}
\begin{center}
\includegraphics[width=6in]{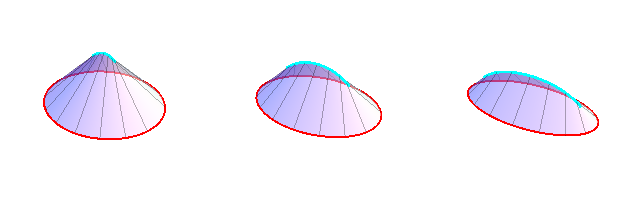}
\caption{
Examples of caustics (top cyan curves) $\caustic^+$ on $\domdy^+[{\cal A}]$ for various shapes of the boundary region ${\cal A}$ (enclosed by the bottom red curves).  
For nearly-circular ellipse, $\caustic^+$ is very short (and would degenerate to a point for circular region); as the ellipse gets more elongated, the caustic arc stretches.
}
\label{f:caustics}
\end{center}
\end{figure}
The behaviour of caustics for ellipsoidal region ${\cal A}$ is illustrated in \fig{f:caustics}.

Let us make a few remarks about the location and structure of caustics.
For smooth $\entsurf$, caustics don't extend all the way down to $t=0$ at which ${\cal A}$ is located.  They only extend to minimal time given by the smallest radius of curvature of $\entsurf$. 
Moreover, they should form a connected tree-like structure.  The maximal time to which they can reach is bounded from above by the maximal extent of ${\cal A}$; however for elongated regions this is a very weak bound.  For example, for rectangular regions, the maximal time reached by $\caustic^+$ is given by half the width, rather than the length, of the rectangle. This in turn puts a bound on the bulk extent of $\cwedge$.

%

\providecommand{\href}[2]{#2}\begingroup\raggedright\endgroup

\end{document}